\documentclass[journal,twocolumn]{IEEEtran}
\usepackage{algorithm}
\pdfoutput=1
\usepackage{algpseudocode}%
\usepackage[utf8]{inputenc}
\usepackage[T1]{fontenc}
\usepackage{url}
\usepackage{ifthen}
\usepackage{cite}
\usepackage[cmex10]{amsmath}
\usepackage{authblk}
\usepackage{ amssymb ,centernot,bbm}
\usepackage[final]{graphicx}
\usepackage{subcaption}
\usepackage{diagbox}

\usepackage{textcomp}

\usepackage{tikz}
	\usetikzlibrary{decorations.pathreplacing,intersections}
\usepackage{cite}
\usepackage{hyperref}
\usepackage{mathtools}

\DeclareMathOperator*{\argmin}{arg\,min}

\usepackage{amsthm}
\newtheorem{theorem}{Theorem}
\newtheorem{lemma}{Lemma}
\newtheorem{proposition}{Proposition}
\newtheorem{corollary}{Corollary}

\theoremstyle{remark}
\newtheorem{rem}{Remark}

\title{The Optimal Memory-Rate Trade-off for the Non-uniform Centralized Caching Problem with Two Files under Uncoded Placement}

\author{Saeid~Sahraei,~\IEEEmembership{Member,~IEEE,}
        Pierre~Quinton,~\IEEEmembership{Student Member,~IEEE,}
        and~Michael~Gastpar,~\IEEEmembership{Fellow,~IEEE}
\thanks{This work was supported in part by the Swiss National Science Foundation under Grants 169294 and 178309. This paper was partially presented at the International Zurich Seminar on Information and Communication, 2018.}
\thanks{Saeid Sahraei is with the Department
of Electrical Engineering, University of Southern California, Los Angeles,
CA 90089, USA (e-mail: ss\_805@usc.edu).}
\thanks{Pierre Quinton and Michael Gastpar are with the School of Computer and Communication Sciences, \'Ecole Polytechnique F\'ed\'erale de Lausanne (EPFL), Lausanne, Vaud 1015, Switzerland (e-mail: pierre.quinton@epfl.ch, michael.gastpar@epfl.ch).}
\thanks{Copyright (c) 2019 IEEE. Personal use of this material is permitted.  However, permission to use this material for any other purposes must be obtained from the IEEE by sending a request to pubs-permissions@ieee.org.
}}

\begin{document}

\maketitle 

\begin{abstract}
A new scheme for the problem of centralized coded caching with non-uniform demands is proposed. The distinguishing feature of the proposed placement strategy is that it admits equal sub-packetization for all files while allowing the users to allocate more cache to the files which are more popular. This creates natural broadcasting opportunities in the delivery phase which are simultaneously helpful for the users who have requested files of different popularities. For the case of two files, we propose a new delivery strategy based on interference alignment which enables each user to decode his desired file following a two-layer peeling decoder. {Furthermore, we extend the existing converse bounds for uniform demands under uncoded placement to the nonuniform case. To accomplish this, we construct $N!$ auxiliary users, corresponding to all permutations of the $N$ files, each caching carefully selected sub-packets of the files. Each auxiliary user provides a different converse bound. The overall converse bound is the maximum of all these $N!$ bounds.} We prove that our achievable delivery rate for the case of two files  meets this converse, thereby establishing the optimal expected memory-rate trade-off for the case of $K$ users and two files with arbitrary popularities under uncoded placement. 
\end{abstract}


\section{Introduction} 
\IEEEPARstart{W}{ireless} traffic has been dramatically increasing in recent years, mainly due to the increasing popularity of video streaming services. Caching is a mechanism for Content Distribution Networks (CDNs) to cope with this increasing demand by placing the contents closer to the users during off-peak hours. The attractive possibility of replacing expensive bandwidth with cheap memories has caused an outburst of research in the recent years \cite{maddah2014fundamental,yu2017characterizing,yu2017exact,pedarsani2016online,karamchandani2016hierarchical,ji2016fundamental, ghasemi2017improved, golrezaei2012femtocaching,wang2016new}. 
Coded caching \cite{maddah2014fundamental} is a canonical formulation of a two-stage communication problem between a server and many clients which are connected to the server via a shared broadcast channel. The two stages consist of filling in the caches of the users during off-peak hours and transmitting the desired data to them at their request, typically during the peak hours. The local caches of the users act as side information for an instance of the index coding problem where different users may have different demands. Logically, if a content is more likely to be requested, it is desirable to cache more of it during the first stage. Furthermore, by diversifying the contents cached at different users, broadcasting opportunities can be created which are simultaneously helpful for several users \cite{maddah2014fundamental}.

In general, there exists a trade-off between the amount of cache that each user has access to and the delivery rate at the second stage. Significant progress has been made towards characterizing this trade-off for worst case and average case demands under uniform file popularities \cite{maddah2014fundamental, maddah2015decentralized,yu2017exact,gomez2016fundamental,yu2017characterizing}. The optimal memory-rate region under uncoded placement is known \cite{yu2017exact,wan2016optimality}, and  the general optimal memory-rate region has been characterized within a factor of 2  \cite{yu2017characterizing}. Furthermore, many achievability results have been proposed based on coded placement \cite{gomez2016fundamental,chen2014fundamental,amiri2016fundamental}. {Some of these schemes outperform the optimal caching scheme under uncoded placement \cite{yu2017exact}}, establishing that uncoded placement is in general sub-optimal. 

By contrast, the coded caching problem with non-uniform file popularities, an arguably more realistic model, has remained largely open. The existing achievability schemes are generally speaking straightforward generalizations of the caching schemes that are specifically tailored to the uniform case. Here we briefly review some of these works.

\subsection{Related Work}
The main body of work on non-uniform coded caching has been concentrated around the decentralized paradigm where there is no coordination among different users \cite{maddah2015decentralized}. The core idea here is to partition the files into $L$ groups where the files within each group have similar popularities \cite{niesen2017coded}. Within each group, one performs the decentralized coded caching strategy of \cite{maddah2015decentralized} as if all the files had the same probability of being requested. In the delivery phase, coding opportunities among different groups are ignored and as a result, the total delivery rate is  the sum of the delivery rates for the $L$ partitions. It was subsequently suggested to use $L=2$ groups \cite{ji2014average,ji2017order,zhang2018coded} and  to allocate no cache at all to the group which contains the least popular files. This simple scheme was proven to be within a multiplicative and additive gap of optimal for arbitrary file popularities \cite{zhang2018coded}.

The problem of {\it centralized} coded caching with non-uniform demands has also been extensively studied \cite{jin2017structural,daniel2017optimization,saberali2018full,ozfatura2018uncoded,ding2017improved}. Here, in order to create coding opportunities among files with varying popularities, a different approach has been taken. Each file is partitioned into $2^K$ subfiles corresponding to all possible ways that a given subfile can be shared among $K$ users. This creates coding opportunities among subfiles that are shared among equal number of users, even if they belong to files with different popularities. The delivery rate can be minimized by solving an optimization problem that decides what portion of each file must be shared among $i$ users, for any $i\in [0:K]$. {It was proven in \cite{saberali2018full} that if the size of the cache belongs to a set of base-cases ${\cal M}$ of size $NK+1$, the best approach is to allocate no cache at all to the least popular files while treating the other files as if they were equally probable of being requested. Memory-sharing among such points must be performed if the cache size is not a member of ${\cal M}$. The set of base-cases depends on the popularities of the files, the number of users and the number of files, and can be computed via an efficient algorithm \cite{saberali2018full}.}

{The graph-based delivery strategies that are predominantly used in this line of research \cite{ji2017order,saberali2018full} are inherently limited, in that they do not capture the algebraic properties of summation over finite fields. For instance, one can easily construct examples where the chromatic-number index coding scheme in \cite{ji2017order} is sub-optimal, even for uniform file popularities. Suppose we have only one file $A=\{A_1,A_2,A_3\}$ and 3 users each with a cache of size $1/3$ file. Assume user $i$ caches $A_i$. In this case, the optimal delivery rate is $2/3$ but the clique-cover (or chromatic-number) approach in \cite{ji2017order} only provides a delivery rate of 1.  This is due to the fact that from $A_1 \oplus A_2$ and $A_2 \oplus A_3$ one can recover $A_1 \oplus A_3$, a property that the graph-based model in \cite{ji2017order} fails to reflect. This issue was addressed in a systematic manner by Yu et. al. in \cite{yu2017exact} which introduced the concept of leaders. Our delivery scheme in this paper provides an alternative mechanism  for overcoming this limitation, which transcends the uniform file popularities and can be applied to nonuniform demands. This is accomplished via interference alignment as outlined in the proof of achievability of Lemma \ref{lemma:lower_compress} in Section \ref{sec:delivery}. }

In \cite{ding2017improved} it was proven that a slightly modified version of the decentralized scheme in \cite{ji2014average} is optimal for the centralized caching problem when we have only two users. In \cite{ozfatura2018uncoded}, a centralized caching strategy was proposed for the case where the number of users $K$ is prime and the case where $K$ divides ${K\choose t}$,  where ${K\choose t}$ is the subpacketization of each file. The placement scheme allows for equal subpacketization of all the files while more stringent requirements are imposed for caching subfiles of less popular files. This concept is closely related to what was presented  in \cite{quinton2018novel} which serves as the placement scheme for the current paper.

\subsection{Our Contributions}
In this paper, we first propose a centralized placement strategy for an arbitrary number of users and files, which allows for equal subpacketization of all the files while allocating less cache to the files which are less likely to be requested. This creates natural coding opportunities in the delivery phase among all the files regardless of their popularities.  Next, we propose a delivery strategy for the case of two files and an arbitrary number of users. { This delivery strategy consists of two phases. First, each file is compressed down to its entropy conditioned on the side information available at the users who have requested it. Simultaneously, this encoding aims at {\it aligning} the subfiles which are unknown to the users who have {\it not} requested them. In the second phase of the delivery strategy, the two encoded files are further encoded with an MDS code and broadcast to the users. Each user will be able to decode his desired file following a two-layer peeling decoder. By extending the converse bound for uncoded placement first proposed in \cite{yu2017exact} to the  non-uniform case, we prove that our joint placement and delivery strategy is optimal for two files with arbitrary popularities under uncoded placement. To summarize, our main contributions are the following:
\begin{itemize}
\item A new placement strategy is developed for non-uniform caching with $K$ users and $N$ files (Section \ref{sec:placement}). This scheme allows for equal sub-packetization of every file, while allocating more cache to files that are more popular. A simple modification of the proposed scheme can be applied to user-dependent file popularities. More broadly, the proposed multiset indexing approach to subpacketization can be expected to find applications in other coding problems of combinatorial nature with heterogeneous objects, such as Coded Data shuffling \cite{lee2018speeding}, Coded Map-Reduce \cite{li2015coded}, and Fractional Repetition codes \cite{el2010fractional} for Distributed Storage. 
\item An extension of the converse bound under uncoded placement first proposed in \cite{yu2017exact} to non-uniform caching with $K$ users and $N$ files is established (Section \ref{sec:converse}).
\item A new delivery strategy is presented for the case of two files which relies on source coding and interference alignment (Section \ref{sec:delivery}). The achievable expected delivery rate meets the extended converse bound under uncoded placement, hence establishing the optimal memory-rate tradeoff for non-uniform demands for the case of two files. If each file has probability $1/2$, this approach leads  to an alternative delivery strategy for uniform caching of two files, which can be of independent interest.
\end{itemize} 
}

The rest of the paper is organized as follows. We introduce the notation used throughout the paper and the formal problem statement in Sections \ref{sec:notation} and \ref{sec:model}. In Section \ref{sec:example} we will explain the main ideas behind our caching strategy via a case study. The general placement and delivery strategy are presented in Sections \ref{sec:placement} and \ref{sec:delivery}. We will then propose our converse bound under uncoded placement in Section \ref{sec:converse}. In Section \ref{sec:optimality} we will prove that our proposed caching strategy is optimal for the case of two files. Finally, in Section \ref{sec:numerical}, we will provide numerical results, and conclude the paper in Section \ref{sec:conclusion}.

\section{Notation}
\label{sec:notation}
For two integers $a,b$ define ${a \choose b} = 0$ if $b<0$ or $b>a$. For $n+1$ non-negative integers $a,b_1,\dots,b_n$ that satisfy $\sum_{i=1}^N b_i = a$, define 
\begin{eqnarray}
{a \choose b_1,\dots,b_n} &=& {a \choose b_1}{a - b_1 \choose b_2}\cdots { a -\sum_{i =1}^{N-1} b_{i}\choose b_n}\nonumber\\ &=& \frac{a!}{b_1!\cdots b_n!}.
\end{eqnarray}
For a positive integer $a$ define $[a] = \{1,\dots,a\}$. For two integers $a,b$ define $[a:b] = \{a,a+1,\dots,b\}$. {For two column vectors ${\bf u}$ and ${\bf v}$ denote their vertical concatenation by $[{\bf u}; {\bf v}]$. For a real number $a$, define $\lfloor a\rfloor$ as the largest integer no greater than $a$. Similarly, define $\lceil a\rceil$ as the smallest integer no less than $a$.}
For $q\in \mathbb{R}^+$  and a discrete random variable $X$ with support ${\cal X}$ define $H_q(X)$ as the entropy of $X$ in base $q$:
\begin{eqnarray}
H_q(X) =- \sum_{x\in {\cal X}} \mathbb{P}(X =x)\log_q\mathbb{P}(X =x).
\end{eqnarray}
Suppose we have a function $f(\cdot): {\cal D}\rightarrow \mathbb{R}$ where ${\cal D}$ is a discrete set of points in $\mathbb{R}^n$. Let ${\cal T}$ be the convex hull of ${\cal D}$. Define
\begin{eqnarray}
g({\bf t}) &=& {\cal L}_{{\bf r}\rightarrow {\bf t}} f({\bf r})\nonumber\\
g(\cdot)&:& {\cal T}\rightarrow \mathbb{R}
 \end{eqnarray}
 as the lower convex envelope of $f(\cdot)$ evaluated at point ${\bf t}\in {\cal T}$.  

\section{Model Description and Problem Statement}
\label{sec:model}
\begin{figure}
\centering
\begin{tikzpicture}[line width = 0.5mm]
\draw (0,0) rectangle (1,1);
\draw (1.3,0) rectangle (2.3,1);
\draw (2.6,0) rectangle (3.6,1);
\draw (3.9,0) rectangle (4.9,1);

\draw(0.5,0.5) node{$Z_1$};
\draw(1.8,0.5) node{$\dots$};
\draw(3.1,0.5) node{$\dots$};
\draw(4.4,0.5) node{$Z_K$};

\draw[<-] (0.5,1) -- (0.5,2.8);
\draw[<-] (1.8,1) -- (1.8,2.8);
\draw[<-] (3.1,1) -- (3.1,2.8);
\draw [<-](4.4,1) -- (4.4,2.8);
\draw(0.5,2.8) -- (4.4,2.8);

\draw (0.9,5) rectangle (1.9,6);
\draw (2,5) rectangle (3,6);
\draw (3.1,5) rectangle (4.1,6);
\draw[dashed] (0.7,4.8) rectangle (4.3,6.2);

\draw(1.4,5.5) node{$W_1$};
\draw(2.5,5.5) node{$\dots$};
\draw(3.6,5.5) node{$W_N$};
\draw(5,5.5) node{Server};
\draw(5.6,0.5) node{Users};

\draw[->] (2.5,4.8) -- (2.5,2.8);

\end{tikzpicture}
\caption{An illustration of the caching network. A server is connected to $K$ users via a shared broadcast link. Each user has a cache of size $MF$ symbols where he can store an arbitrary function of the files  $W_1,\dots,W_N$.}
\label{fig:setup}
\end{figure}
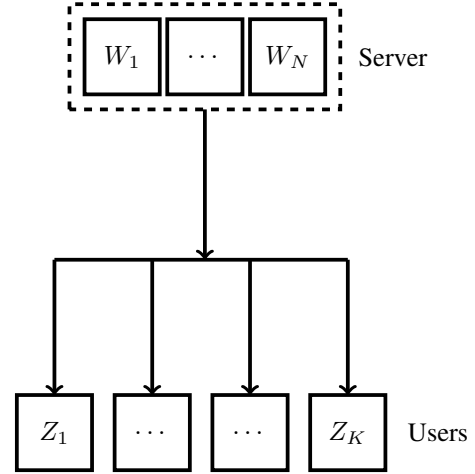

We follow the canonical model in \cite{maddah2014fundamental} except here we concentrate on the expected delivery rate as opposed to the worst case delivery rate. For the sake of completeness, we repeat the model description here. We have a network consisting of $K$ users that are connected to a server through a shared broadcast link. The server has access to $N$ files $W_1,\dots,W_N$ each of size $F$ symbols over a sufficiently large field $\mathbb{F}_q$. Therefore, $H_q(W_i)\le F$. Each user has a cache of size $M$ symbols over ${\mathbb{F}}_q$. An illustration of the network has been provided in Figure \ref{fig:setup}. The communication between the server and the users takes place in two phases, placement and delivery.

In the placement phase, each user stores some function of all the files  $Z_i = f_i(W_1,\dots,W_N), i\in[K]$ in his local cache. Therefore, for a fixed (normalized) memory size $M$, a placement strategy ${\cal M} $ consists of $K$ placement functions $Z_i = f_i(W_1,\dots,W_N), i\in[K]$ such that  $H_q(Z_i)\le MF$ for all $i\in [K]$. After the placement phase, each user requests one file from the server. We represent the request of the $i$'th user with $d_i\in[N]$ which is drawn from a known probability distribution ${\bf p}$. Furthermore, the requests of all the users are independent and identically distributed. After receiving the request vector ${\bf d}$, the server transmits a delivery message $X_{{\bf d},{\cal M},F}$ through the broadcast link to all the users. User $i$ then computes a function $\hat{W}_{d_i} = g_i(X_{{\bf d},{\cal M},F},Z_i,{\bf d})$ in order to estimate $W_{d_i}$. For a fixed placement strategy ${\cal M}$, fixed file size $F$, and fixed request vector ${\bf d}$ we say that a delivery rate of $R_{{\bf d},{\cal M},F}$ is achievable if a delivery message $X_{{\bf d},{\cal M},F}$ and decoding functions $g_i(\cdot), i\in[K]$ exist such that 
\begin{eqnarray}
\mathbb{P}(g_i(X_{{\bf d},{\cal M},F},Z_i,{\bf d}) \neq W_{d_i}) = 0, \; \forall i\in [N],
\end{eqnarray} 
and 
\begin{eqnarray}
H_q(X_{{\bf d},{\cal M},F}) \le R_{{\bf d},{\cal M},F}F.
\end{eqnarray}
For a fixed placement strategy ${\cal M}$,  we say that an expected delivery rate $\bar{R}_{\cal M}$ is achievable if there exists a sequence of achievable delivery rates $\{R_{{\bf d},{\cal M},F}| {\bf d}\in [N]^K, F\in\mathbb{ N}\}$  such that 
\begin{eqnarray}
 \limsup_{F\rightarrow \infty}\mathbb{E}_{\bf d}R_{{\bf d},{\cal M},F}\le \bar{R}_{\cal M}.
\end{eqnarray}
Finally, for a memory of size $M$, we say that an expected delivery rate $\bar{R}$ is achievable if there exists a placement strategy ${\cal M}= (Z_1,\dots,Z_K)$ with $H_q(Z_i)\le MF$ for all $i\in [K]$, for which an expected delivery rate of $\bar{R}_{\cal M}\le\bar{R}$ is achievable. \\
Our goal in this paper is to characterize the minimum expected delivery rate for all $M$ under the restriction of uncoded placement. In other words, the placement functions must be of the form 
\begin{eqnarray}
Z_i &=& f_i(W_1,\dots,W_N)\nonumber\\ &=& ({W_{1}|_{A_1},\dots,W_{N}|_{A_N}}) \mbox{ for all } i\in[K],
\end{eqnarray}
where $A_j\subseteq[F]$ for all $j\in[N]$, and {$W_j|_{A_j}$} refers to the subset of symbols of the file $W_j$ which are indexed in the set $A_j$. 

\section{Motivating Example: The Case of Four Users and Two Files}
\label{sec:example}
Consider the caching problem with two files $W_1$ and $W_2$ and $K =4$ users. Assume the probability of requesting $W_2$ is lower than the probability of requesting $W_1$. {In this section we will demonstrate how to find the optimal expected delivery rate for any memory size for this particular choice of parameters, while explaining the main principles behind our joint placement and delivery strategy. We start by fixing two integers $r_1,r_2$ such that $0\le r_2 \le r_1\le K$. As we will see soon, any choice of $(r_1,r_2)$ corresponds to a particular $(M_1,M_2)$ where $M_i$ is the amount of cache that each user allocates to file $W_i$, {normalized by the size of one file}. For the sake of brevity, we will explain our strategy only for $(r_1,r_2) = (2,1)$. The delivery rate for other possible choices of $(r_1,r_2)$ will be summarized at the end of this section. Next, we will characterize the entire $(M_1,M_2,R)$ region that can be achieved by our algorithm. Finally, we will illustrate how to find the optimal choice of $(M_1,M_2)$ for a particular cache size $M$.}\\
\indent Define the parameter $S = {K \choose r_1}{r_1 \choose r_2}$. We divide each of the two files into $S$ subfiles and index them as $W_{1,\tau_1,\tau_2}$ and $W_{2,\tau_1,\tau_2}$ such that $\tau_2\subseteq\tau_1\subseteq [K]$ and $|\tau_1| =r_1, |\tau_2| =r_2$. In this example, $S = {4 \choose 2}{2 \choose 1} = 12.$ The 12 subfiles of $W_i$ are then denoted by $W_{i,\{1,2\},\{1\}},$ $W_{i,\{1,2\},\{2\}},$ $W_{i,\{1,3\},\{1\}},$ $W_{i,\{1,3\},\{3\}},$ $W_{i,\{1,4\},\{1\}},$ $W_{i,\{1,4\},\{4\}},$ $ W_{i,\{2,3\},\{2\}},$ $W_{i,\{2,3\},\{3\}},$ $W_{i,\{2,4\},\{2\}},$ $W_{i,\{2,4\},\{4\}},$ $W_{i,\{3,4\},\{3\}},$ $W_{i,\{3,4\},\{4\}}$. In our placement strategy, user $j$ stores the subfiles $W_{1,\tau_1,\tau_2}$ for which $j\in \tau_1,$ as well as the subfiles $W_{2,\tau_1,\tau_2}$ for which $j\in \tau_2$. Since $\tau_2\subseteq\tau_1$, the users naturally store fewer subfiles of $W_2$ than $W_1.$ In our running example, each user stores six subfiles of $W_1$ but only three subfiles of $W_2.$ The cache contents of each user has been summarized in Table~\ref{tab:placement}.
\begin{table}
\begin{center}
\begin{tabular}{|c|c|}
\hline
$Z_1$& $Z_2$\\
\hline 
$W_{1,\{1,2\},\{1\}}, W_{1,\{1,2\},\{2\}}$ & $W_{1,\{1,2\},\{2\}}, W_{1,\{1,2\},\{1\}}$   \\
$W_{1,\{1,3\},\{1\}}, W_{1,\{1,3\},\{3\}}$ & $W_{1,\{2,3\},\{2\}}, W_{1,\{2,3\},\{3\}}$ \\
$W_{1,\{1,4\},\{1\}}, W_{1,\{1,4\},\{4\}}$ & $W_{1,\{2,4\},\{2\}}, W_{1,\{2,4\},\{4\}}$\\
\hline
$W_{2,\{1,2\},\{1\}}$ & $W_{2,\{1,2\},\{2\}}$   \\
$W_{2,\{1,3\},\{1\}}$ & $W_{2,\{2,3\},\{2\}}$   \\
$W_{2,\{1,4\},\{1\}}$ & $W_{2,\{2,4\},\{2\}}$  \\
\hline
$Z_3$ & $Z_4$\\
\hline
$W_{1,\{1,3\},\{3\}}, W_{1,\{1,3\},\{1\}}$ & $W_{1,\{1,4\},\{4\}}, W_{1,\{1,4\},\{1\}}$ \\
 $W_{1,\{2,3\},\{3\}}, W_{1,\{2,3\},\{2\}}$ & $W_{1,\{2,4\},\{4\}}, W_{1,\{2,4\},\{2\}}$ \\
 $W_{1,\{3,4\},\{3\}}, W_{1,\{3,4\},\{4\}}$ & $W_{1,\{3,4\},\{4\}}, W_{1,\{3,4\},\{3\}}$ \\
\hline
$W_{2,\{1,3\},\{3\}}$ & $W_{2,\{1,4\},\{4\}}$\\
$W_{2,\{2,3\},\{3\}}$ & $W_{2,\{2,4\},\{4\}}$\\
$W_{2,\{3,4\},\{3\}}$ & $W_{2,\{3,4\},\{4\}}$ \\
\hline
\end{tabular}
\end{center}
\caption{The proposed placement scheme for $N =2 $ , $K=4$, $(r_1,r_2) = (2,1)$.}
\label{tab:placement}
\end{table}

Note that this placement scheme results in a memory of size $M =\frac{3}{4}$. As we will see soon, the memory size is in general $  M = \sum_{i=1}^N M_i $ where $M_i = \frac{r_i}{K}$ is the amount of cache dedicated by each user to file $W_i$. It is important to note that despite the fact that each user has allocated more cache to file $W_1$, all the subfiles are of equal size. This is a key property of the proposed placement scheme which allows us to efficiently transmit messages in the delivery phase which are simultaneously helpful for users who have requested files of different popularities. 

Let us now turn to the delivery phase. To make matters concrete, let us suppose that the first three users have demanded $W_1$ and the last user is interested in $W_2$. Therefore, our demand vector is $d = (1,1,1,2)$. We define $\Omega_i$ as the subset of users who have requested file $W_i$. In this case, $\Omega_1 = \{1,2,3\}$ and $\Omega_2 = \{4\}$.

For the delivery phase, we construct a compressed description $W_i^*$ for each file $W_i.$ 
For users in $\Omega_i$ recovering $W^*_i$ implies recovering $W_i$, that is, $H_q(W_i|W_i^*,Z_j) = 0$. Moreover, among all $W^*_i$ that satisfy this property, our particular construction minimizes both $\max_{j\in\Omega_{ i}} H_q(W_i^*|Z_j)$ and $\max_{j\notin\Omega_{ i}}H_q(W^*_i|Z_j)$ at the same time. The general construction of $W^*_i$ is presented in Section~\ref{sec:delivery}, along with proofs of its properties.
For the example at hand, our construction specializes to
\begin{eqnarray}
W^{*}_{1} = \left[W^{*}_{1,\{4\},\{4\}} ; W^{*}_{1,\{4\},\{\}};W^{*}_{1,\{\},\{\}}\right],
\end{eqnarray}
where 
\begin{eqnarray}
W^{*}_{1,\{4\},\{4\}} &=& \left[W^{*(1)}_{1,\{4\},\{4\}}; W^{*(2)}_{1,\{4\},\{4\}}\right],\nonumber\\
W^{*}_{1,\{4\},\{\}} &=& \left[W^{*(1)}_{1,\{4\},\{\}}; W^{*(2)}_{1,\{4\},\{\}}\right],\nonumber\\
W^{*}_{1,\{\},\{\}} &=& \left[W^{*(1)}_{1,\{\},\{\}}; W^{*(2)}_{1,\{\},\{\}}\right],
\end{eqnarray}
and
\begin{align}
W^{*(1)}_{1,\{4\},\{4\}} &=  W_{1,\{1,4\},\{4\}} + W_{1,\{2,4\},\{4\}}+ W_{1,\{3,4\},\{4\}}, \nonumber\\
W^{*(2)}_{1,\{4\},\{4\}} &=    W_{1,\{1,4\},\{4\}} + 2W_{1,\{2,4\},\{4\}}+ 3W_{1,\{3,4\},\{4\}},\nonumber\\
W^{*(1)}_{1,\{4\},\{\}} &= W_{1,\{1,4\},\{1\}} + W_{1,\{2,4\},\{2\}}+ W_{1,\{3,4\},\{3\}},\nonumber\\
W^{*(2)}_{1,\{4\},\{\}} &= W_{1,\{1,4\},\{1\}} + 2W_{1,\{2,4\},\{2\}}+ 3W_{1,\{3,4\},\{3\}},\nonumber\\
W^{*(1)}_{1,\{\},\{\}} &=  W_{1,\{1,2\},\{1\}} + W_{1,\{1,2\},\{2\}} + W_{1,\{1,3\},\{1\}} \nonumber
\\&+W_{1,\{1,3\},\{3\}} + W_{1,\{2,3\},\{2\}} + W_{1,\{2,3\},\{3\}}, \nonumber\\
W^{*(2)}_{1,\{\},\{\}} &=   W_{1,\{1,2\},\{1\}} + 2W_{1,\{1,2\},\{2\}} + W_{1,\{1,3\},\{1\}}\nonumber
 \\&+ 2W_{1,\{1,3\},\{3\}} + W_{1,\{2,3\},\{2\}} + 2W_{1,\{2,3\},\{3\}}.
\end{align}
Therefore, the subfiles of $W^*_1$ are $W^{*(1)}_{1,\{4\},\{4\}}, W^{*(2)}_{1,\{4\},\{4\}} ,$ $W^{*(1)}_{1,\{4\},\{\}},  W^{*(2)}_{1,\{4\},\{\}}, W^{*(1)}_{1,\{\},\{\}} , W^{*(2)}_{1,\{\},\{\}}$.  We can represent this in matrix form as follows. 
\normalsize
\begin{align}
&W_1^* = \nonumber\\
\vspace{-5cm}
&\begin{bmatrix}
0&0&0&0&0&1&0&0&0&1&0&1\\
0&0&0&0&0&1&0&0&0&2&0&3\\
0&0&0&0&1&0&0&0&1&0&1&0\\
0&0&0&0&1&0&0&0&2&0&3&0\\
1&1&1&1&0&0&1&1&0&0&0&0\\
1&2&1&2&0&0&1&2&0&0&0&0
\end{bmatrix}
\begin{bmatrix}
W_{1,\{1,2\},\{1\}} \\W_{1,\{1,2\},\{2\}} \\ W_{1,\{1,3\},\{1\}} \\W_{1,\{1,3\},\{3\}} \\ W_{1,\{1,4\},\{1\}} \\W_{1,\{1,4\},\{4\}}  \\W_{1,\{2,3\},\{2\}} \\W_{1,\{2,3\},\{3\}} \\ W_{1,\{2,4\},\{2\}} \\W_{1,\{2,4\},\{4\}} \\W_{1,\{3,4\},\{3\}} \\ W_{1,\{3,4\},\{4\} }
\end{bmatrix}.
\end{align}
 If a user in $\Omega_1$ successfully receives $W_1^*$, he can, with the help of side information already stored in his cache, recover the entire $W_1$. For instance, user $1$ only needs to solve the following set of equations for $W_{1,\{2,3\},\{2\}}, W_{1,\{2,3\},\{3\}},$ $W_{1,\{2,4\},\{2\}},$ $W_{1,\{2,4\},\{4\}} ,W_{1,\{3,4\},\{3\}} ,W_{1,\{3,4\},\{4\}}.$
 {
 \begin{align}
&\begin{bmatrix}
W^{*(1)}_{1,\{4\},\{4\}} \\W^{*(2)}_{1,\{4\},\{4\}}  \\ W^{*(1)}_{1,\{4\},\{\}}\\W^{*(2)}_{1,\{4\},\{\}}\\W^{*(1)}_{1,\{\},\{\}}  \\ W^{*(2)}_{1,\{\},\{\}} 
\end{bmatrix} - \begin{bmatrix}
0&0&0&0&0&1\\
0&0&0&0&0&1\\
0&0&0&0&1&0\\
0&0&0&0&1&0\\
1&1&1&1&0&0\\
1&2&1&2&0&0
\end{bmatrix}
\begin{bmatrix}
W_{1,\{1,2\},\{1\}} \\W_{1,\{1,2\},\{2\}} \\ W_{1,\{1,3\},\{1\}} \\W_{1,\{1,3\},\{3\}} \\W_{1,\{1,4\},\{1\}} \\ W_{1,\{1,4\},\{4\}}
\end{bmatrix}\nonumber\\
&=
\begin{bmatrix}
0&0&0&1&0&1\\
0&0&0&2&0&3\\
0&0&1&0&1&0\\
0&0&2&0&3&0\\
1&1&0&0&0&0\\
1&2&0&0&0&0
\end{bmatrix}
\begin{bmatrix}
W_{1,\{2,3\},\{2\}} \\W_{1,\{2,3\},\{3\}} \\ W_{1,\{2,4\},\{2\}} \\W_{1,\{2,4\},\{4\}} \\W_{1,\{3,4\},\{3\}} \\ W_{1,\{3,4\},\{4\}}
\end{bmatrix}.
\end{align}
This is possible since user 1 knows the left-hand side of the equation, and the matrix on the right hand-side is invertible.} Therefore, our goal boils down to transferring the entire $W_1^*$ to all the users in $\Omega_1$. Following a similar process, we construct the description $W^*_2$ as follows
\begin{align}
W^*_2 &= \left[W^*_{2,\{1\},\{1\}};W^*_{2,\{2\},\{2\}};W^*_{2,\{3\},\{3\}};\right. \nonumber\\
& \left.  W^*_{2,\{1,2\},\{1\}} ; W^*_{2,\{1,2\},\{2\}};W^*_{2,\{1,3\},\{1\}};\right.\nonumber \\
 &\left.W^*_{2,\{1,3\},\{3\}}; W^*_{2,\{2,3\},\{2\}};W^*_{2,\{2,3\},\{3\}} \right],
\end{align}
where
\begin{align}
W^{*}_{2,\{i\},\{i\}} &= W_{2,\{i,4\},\{i\}}\;\; \forall i\in \{1,2,3\},\nonumber\\
W^{*}_{2,\{i,j\},\{i\}} &= W_{2,\{i,j\},\{i\}} \;\; \forall (i,j) \mbox{ s.t. } i,j\in [3], i\neq j.
\end{align}
That is, in this example, $W^*_2$ consists of the subfiles of $W_2$ which are unknown to user $4$. 
Again, transferring the entire $W_2^*$ to user 4, guarantees his successful recovery of $W_2$. \\
\indent To simplify matters, we will require every user in $[K]$ to recover the entire $[W_1^*;W_2^*]$. 
In order to accomplish this, we transmit ${\cal C}[W_1^*;W_2^*]$  over the broadcast link. The matrix ${\cal C}$ here is an MDS matrix of $12$ rows and $15$ columns. The number of rows of this matrix is determined by the maximum number of subfiles of $[W_1^*;W_2^*]$ which are unknown to any given user. In this example, a user in $\Omega_1$ has precisely 12 unknowns in $[W_1^*;W_2^*]$ (6 subfiles of $W_1^*$ and 6 subfiles of $W_2^*$). On the other hand,  a user in $\Omega_2$ knows 4 out of the 6 subfiles of $W_1^*$. Therefore, a total of 11 subfiles of $[W_1^*;W_2^*]$ are unknown to him. Hence, the matrix ${\cal C}$ must have 12 rows. Once a user in $[K]$ receives ${\cal C}[W_1^*;W_1^*]$, he can remove the columns of ${\cal C}$ which correspond to the subfiles he already knows. The resulting matrix will be square (or overdetermined) which is invertible owing to the MDS structure of ${\cal C}$. This will allow every user to decode $[W_1^*;W_2^*]$. Subsequently, each user in $\Omega_i$ can proceed to decode $W_i$ with the help of his side information. {Recall that we started by dividing each file into $12$ subfiles, and the delivery message consists of 12 linear combinations of such subfiles. Therefore, the delivery rate for this particular request vector is $R = 1$.}\\
\indent As we will see in the next section, the delivery rate of our strategy only depends on the request vector ${\bf d}$ through ${\cal N} = Range({\bf d})$, the set of indices of all the files that have been requested at least once.  Therefore, we showed that with $N = 2, K = 4$, $(r_1,r_2) = (2,1)$, and assuming ${\cal N} = \{1,2\}$, we can achieve a delivery rate of $R = 1$. We can perform the same process for every choice of $(r_1,r_2)\in\mathbb{Z}^2$ that satisfies $0\le r_2\le r_1\le K$. The result is summarized in Table \ref{tab:allr}. Note that if ${\cal N} = \{i\}$, the delivery rate is simply $1 - \frac{r_i}{K}$.  

\begin{figure}
\centering
\includegraphics[scale=0.31]{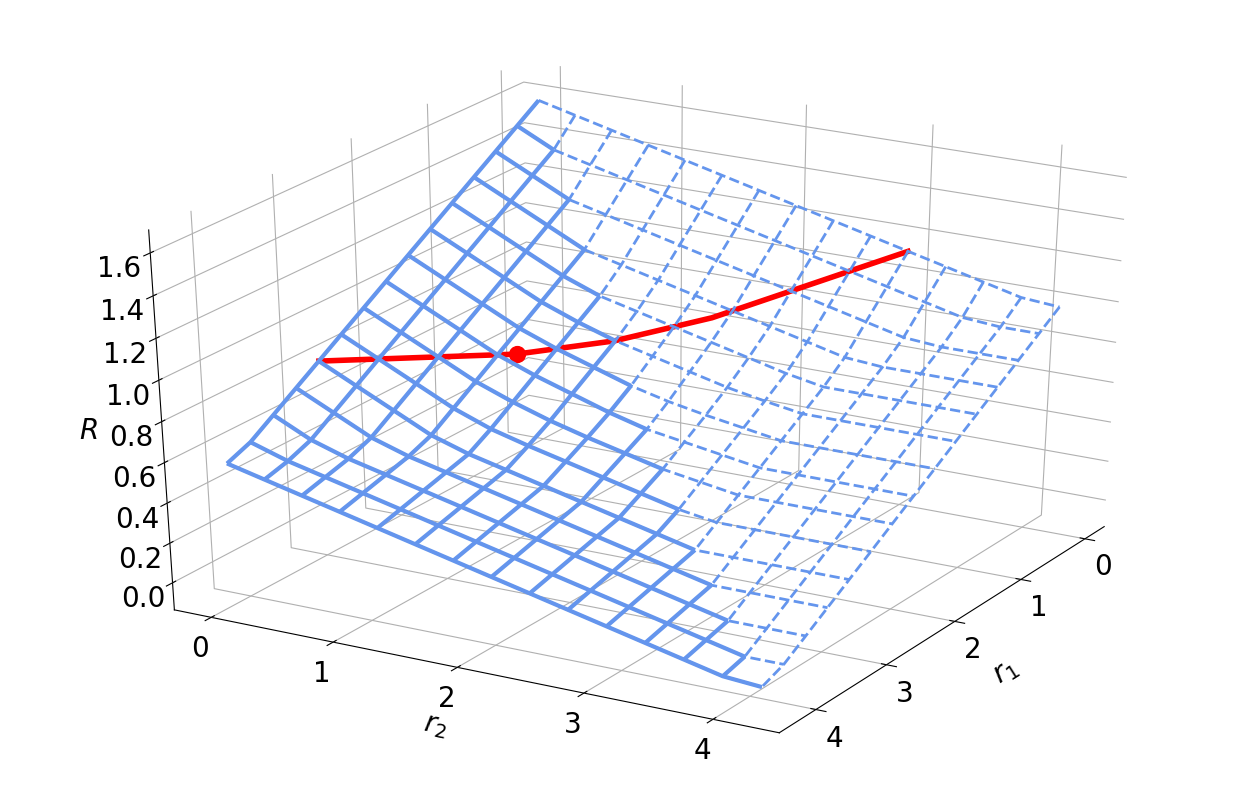}
  \caption{The expected delivery rate for the caching problem with 4 users and 2 files versus $(r_1,r_2)$. The probabilities of the two files are $0.8$ and $0.2$ respectively. The thick red curve determines the set of $(r_1,r_2)$ which results in a cache of size $M = 3/4$. The red circle on the curve is the minimizer of the red curve, which provides the optimal delivery rate under uncoded placement for the given $(K,N,{\bf p},M)$. }
 \label{fiig:r1r2_3d}
  \end{figure}

\begin{table}[H]
\begin{center}
\addtolength{\tabcolsep}{-2pt}
\begin{tabular}{|c|c|c|c|c|c|}
\hline
\diagbox[width=6em]{\hphantom{spac}${\cal N}$}{$(r_1,r_2)$}&$(0,0)$&$(1,0)$&$(1,1)$&$(2,0)$&$(2,1)$\\
\hline
$\{1,2\} $&$2$&$7/4$&$5/4$&$3/2$&$1$\\
$\{1\}$ &$1$&$3/4$&$3/4$&$1/2$&$1/2$\\
$\{2\}$ &$1$&$1$&$3/4$&$1$&$3/4$\\
\hline
\diagbox[width=6em]{\hphantom{spac}${\cal N}$}{$(r_1,r_2)$}&$(2,2)$&$(3,0)$&$(3,1)$&$(3,2)$&$(3,3)$\\
\hline
$\{1,2\}$&$2/3$&$5/4$&$3/4$& $1/2$&$1/4$\\
$\{1\}$&$1/2$&$1/4$&$1/4$&$1/4$&$1/4$\\
$\{2\}$&$1/2$&$1$&$3/4$&$1/2$&$1/4$\\
\hline
\diagbox[width=6em]{\hphantom{spac}${\cal N}$}{$(r_1,r_2)$}&$(4,0)$&$(4,1)$&$(4,2)$&$(4,3)$ & $(4,4)$\\
\hline
$\{1,2\}$&$1$&$3/4$&$1/2$&$1/4$&$0$\\
$\{1\}$&$0$&$0$&$0$&$0$&$0$\\
$\{2\}$&$1$&$3/4$&$1/2$&$1/4$&$0$\\
\hline
\end{tabular}
\end{center}
\caption{The set of delivery rates of our proposed scheme for all possible choices of $0\le r_2\le r_1\le K$ and all possible ${\cal N} \subseteq [N]$. We have $N =2$ and $K =4$. }
\label{tab:allr}
\end{table}
By performing memory-sharing among all such points, we are able to achieve the lower convex envelope of the points in Table \ref{tab:allr}. The expected delivery rate {as a function of $(r_1,r_2)$} for a probability distribution of $(p_1,p_2) = (0.8,0.2)$ has been plotted in Figure \ref{fiig:r1r2_3d}. Note that the dotted half of the figure where $r_2 > r_1$  would correspond to switching the roles of the two files $W_1$ and $W_2$ and allocating more cache to the less popular file. The next question is how to find the best delivery rate for a particular cache size $M.$ For this, we first have to restrict Figure \ref{fiig:r1r2_3d} to the trajectory $r_1 + r_2 = MK$. As an example, we have plotted the thick red curve on the figure which corresponds to $r_1 + r_2 = 3$ (or $M=3/4$). In order to find the best caching strategy for a cache size of $M=3/4$, we need to choose the global minimum of this red curve. This can be done efficiently due to the convexity of the curve, and as we will see in Section \ref{sec:binary_search}, can be even performed via binary search over the set of break points of the curve. As marked on the figure with a red circle, for this particular example with $K=4, N=2,(p_1,p_2) = (0.8,0.2), M =3/4$, the expected delivery rate is $0.79$ which can be achieved by allocating a cache of size $M_1 = r_1/K = 0.5$ to file $W_1$ and $M_2 = r_2/K= 0.25$ to file $W_2$. Theorem \ref{thm:optimality} from Section \ref{sec:optimality} will tell us that under the restriction of uncoded placement, this is the best expected delivery rate that one can achieve for the given $(K,N,{\bf p},M)$.

\section{The Placement Strategy}
\label{sec:placement}
\begin{figure}
\centering
\begin{tikzpicture}[line width = 0.5mm]
\draw (0,0) rectangle (7,5);
 \draw[black] (2.5,2.5) circle (2.3);
  \draw[black] (3,2) circle (1.3);
  \Large
  \draw(6,4) node{$[K]$};
    \draw(2.5,4.4) node{$\tau_1$};
        \draw(3,2.7) node{$\tau_2$};
         \normalsize
         \draw(2.5,3.7) node{Cache $W_{1,\tau_1,*}$};         
        \small
  \draw(3,1.8) node{Cache $W_{2,\tau_1,\tau_2}$};
 \normalsize
\end{tikzpicture}
\caption{Van Diagram of the placement strategy for the case of two files. Users whose indices appear in $\tau_1$ cache $W_{1,\tau_1,\tau_2}$ for all $\tau_2\subseteq\tau_1,\;\; |\tau_2| = r_2$. Users whose indices appear in $\tau_2$ cache $W_{2,\tau_1,\tau_2}$.}
\label{fig:van}
\end{figure}
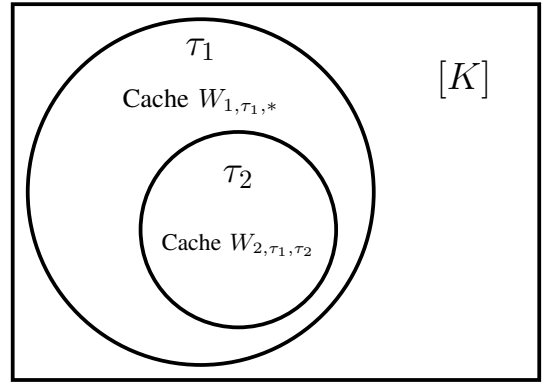

In this section we describe our general placement strategy. Note that our placement strategy can be applied to an arbitrary number of files and users and can even be adapted to user-specific file popularities (see Remark \ref{rem:userspecific}). Without loss of generality, suppose that the files are indexed in decreasing order of their popularity. In other words, file $W_i$ is at least as popular as file $W_{i + 1}$ for all $i\in[N-1]$. The placement strategy begins with selecting integers $r_1,\dots,r_N$ such that  $0\le r_N\le \dots \le r_1\le K$. Each $r_i$ is proportional to the amount of cache that we are willing to allocate to file $W_i$. We divide each file into 
\begin{eqnarray}
S = {K \choose r_N, r_{N-1}- r_N,\dots,r_1 - r_2, K - r_1}
\end{eqnarray}
subfiles of equal size. We label each subfile by $N$ sets ${\tau_1,\dots, \tau_N}$ where $|\tau_j| = r_j$  for $j\in[N]$ and $\tau_j\subseteq \tau_{j-1}$ for $j\in[2:N]$ and $\tau_1 \subseteq[K]$. It should be evident that there are exactly $S$ such subfiles. Next, for file $W_i$, we require each user $k$ to store the subfile $W_{i,\tau_1,\dots,\tau_N}$ if and only if $k\in\tau_i$. This process has been summarized in Algorithm \ref{Alg:placement}, and an illustration for the case of $N =2$ has been provided in Figure \ref{fig:van}. We can compute the amount of cache dedicated by each user to file $i$ as follows
\begin{align}
M_i &= \frac{{K-r_2 \choose r_1 -r_2}\times\dots\times{K - {r_i}\choose r_{i-1} -r_i}{K-1 \choose r_i -1}{r_i \choose r_{i+1}}\times\dots\times {r_{N-1}\choose r_N}}{S}\nonumber\\
&=\frac{{K-r_2 \choose r_1 -r_2}\times\dots\times{K - {r_i}\choose r_{i-1} -r_i}{K-1 \choose r_i -1}{r_i \choose r_{i+1}}\times\dots\times {r_{N-1}\choose r_N}}{{K-r_2 \choose r_1 -r_2}\times\dots\times{K - {r_i}\choose r_{i-1} -r_i}{K\choose r_i }{r_i \choose r_{i+1}}\times\dots\times {r_{N-1}\choose r_N}}\nonumber\\
&=\frac{r_i}{K}.
\end{align}

\noindent This results in a total normalized cache size of

\begin{eqnarray}
M = \sum_{i=1}^N M_i = \frac{\sum_{i=1}^N{r_i}}{K}.
\end{eqnarray}

\begin{rem}
In the special case of $r_1 =\dots = r_N$, all the sets $\tau_i$ will be equal, and can be represented by only one set $\tau$. In this case, our placement phase is equivalent to the uniform placement strategy proposed in \cite{maddah2014fundamental}.
\end{rem}
\begin{rem}
\label{rem:userspecific}
More generally, each user $j$ could choose a permutation $\pi: [N]\rightarrow [N]$ and store file $W_{i,\tau_1,\dots,\tau_N}$ if and only if $j\in \tau_{\pi(i)}$. This would allow different users to have different preferences in terms of the popularities of the files, while still keeping all the cache sizes equal, and maintaining the same sub-packetization for all files. {To provide a simple example, suppose we have two users and two files $W_1$ and $W_2$ with $(r_1,r_2) = (2,1)$ resulting in a cache size of $M = 3/2$. In this case, user $1$ could cache $W_{1,\{1,2\},\{1\}}$ and $W_{1,\{1,2\},\{2\}}$ but only $W_{2,\{1,2\},\{1\}}$. On the other hand, user 2 could cache $W_{2,\{1,2\},\{1\}}$ and $W_{2,\{1,2\},\{2\}}$ but only $W_{1,\{1,2\},\{2\}}$. This caching scheme preserves the property that each file has the same number of subfiles, while allowing each user to give higher priority to a different file.}
\end{rem}
 \begin{algorithm}[H]
\caption[caption]{The Placement Strategy for $N$ files and $K$ users}
\begin{algorithmic}[1]
\Statex {{\bf Input:} $(W_1,\dots,W_N), (r_1,\dots,r_N), K$}
 \Statex {\bf Output: }{The placement contents $(Z_1,\dots,Z_K)$.}
 \Statex{}
 \State $S = {K \choose r_N, r_{N-1}- r_N,\dots,r_1 - r_2, K - r_1}$. 
 \State Break each file $W_i$ into $S$ non-overlapping subfiles of equal size and index them as
\begin{eqnarray*}
W_i = \left\{W_{i,\tau_1,\dots,\tau_N}| \tau_N\subseteq \dots\subseteq \tau_1\subseteq[K], \; |\tau_j| = r_j\;,\; \right. \\ \left. \forall j\in[N]\right\}.
\end{eqnarray*}
\For {$i\in[K]$}
\State $Z_i = \left\{W_{j,\tau_1,\dots,\tau_N}| \mbox{ for all } j\in[N] \mbox{ and all } (\tau_1,\dots,\tau_N) \right.$ $\left.\;\;\;\;\mbox{ such that } i\in \tau_j\right\}.$
\EndFor
\State Return $(Z_1,\dots,Z_K)$. 
\end{algorithmic}
\label{Alg:placement}
\end{algorithm}

\begin{rem}
The combinatorial designs in the Maddah-Ali and Niesen placement strategy \cite{maddah2014fundamental} have recently been used in other closely related fields such as Coded Data Shuffling \cite{lee2018speeding} and Coded Map-Reduce \cite{li2015coded}. Despite being very useful at capturing the symmetric commonalities of different objects, it is not trivial how one can systematically generalize this combinatorial design to heterogeneous networks. We believe that our proposed multiset indexing tool provides a flexible, yet systematic extension of this scheme to asymmetric settings. For instance, consider a uniform caching problem with heterogeneous cache sizes. Let us assume that we have $S$ different cache sizes $M_1 ,\cdots, M_S$. Define $(r_1,\cdots,r_S) = (\frac{KM_1}{N},\cdots,\frac{KM_S}{N})$ and for simplicity assume $r_i\in\mathbb{Z}$. Divide each file $W_i$ into subfiles of equal size $W_{i,\tau_1,\cdots,\tau_S}$ such that $|\tau_j| = r_j$, $\tau_j\subseteq \tau_{j-1}$ and $\tau_1\subseteq[K]$. We could then let the users with cache size $M_j$ store $W_{i,\tau_1,\cdots,\tau_S}$ for all $i\in[N]$ if and only if their indices are in $\tau_j$. As a second example, suppose we want to design a Fractional Repetition code \cite{el2010fractional} for distributed storage where some servers store more data than the others, and play a central role in the data-recovery criterion or the repair process in a distributed storage network. Again, one can systematically design such a storage code with the proposed multiset-indexing framework.
\end{rem}
\section{Delivery Strategy for $N = 2$}
\label{sec:delivery}
Let $\Omega_1,\Omega_2 \subseteq[K]$ represent the subsets of the users that have requested files $W_1$ and $W_2,$ respectively. Therefore $\Omega_1\cap\Omega_2 =\emptyset$ and $\Omega_1\cup\Omega_2 = [K]$. Also define $K_1 = |\Omega_1|$ and $K_2 = |\Omega_2| = K -K_1$. Note that if $K_1 = 0,$ a delivery rate of $R = 1 - \frac{r_2}{K}$ can be trivially achieved. Similarly, if $K_2 = 0$, we can achieve a delivery rate of $R = 1 - \frac{r_1}{K}$. Let us now assume that both files have been requested.

The general idea behind the delivery scheme is as follows. First, we encode each file $W_i$, $i\in[2]$ as $W^*_i$ in such a way that decoding $W^*_i$ provides enough information for each user in $\Omega_i$ to decode $W_i$. In other words, we want that $H_q(W_i|Z_j,W_i^*) =0$ for all $j\in\Omega_i$, $i\in[2]$.  The server transmits sufficient information for all the users in $[K]$ to recover both $W_1^*$ and $W_2^*$. Subsequently, each user in $\Omega_i$ proceeds to recover $W_i$ based on $W_i^*$ and the contents of his cache. Moreover, the goal is for $W_i^*$ to have a significant overlap with the cache of the users outside $\Omega_i$, i.e., $\max_{j\in[K]\backslash\Omega_{i}} H_q(W_i^*|Z_j)$ is as small as possible. The following lemma lays the foundation for our search for the ideal $W^*_1$ and $W^*_2$.  
\begin{lemma}
Suppose $\Omega_1 \neq \emptyset$ and $\Omega_2\neq \emptyset$. For each $i\in[2]$, assume $W_i^*$ satisfies 
\begin{eqnarray}
H_q(W_i|W_i^*,Z_j) = 0,\;\; \forall j\in\Omega_i.
\label{eqn:condition1zero}
\end{eqnarray}
Then $W_i^*$ must satisfy 
\begin{eqnarray}
\max_{m\in\Omega_i}H_q(W_i^*|Z_m)&\ge& \frac{{K-1\choose r_i}}{{K\choose r_i}}F
\label{eqn:condition2mindirect}
\end{eqnarray}
and
\begin{eqnarray}
\max_{\ell\in[K]\backslash\Omega_i}H_q(W_i^*|Z_\ell)&\ge& \frac{{K-2\choose r_i}}{{K\choose r_i}}F.
\label{eqn:condition2mincross}
\end{eqnarray}
Furthermore, there exist $W^*_1$ and $W^*_2$ that satisfy Equations \eqref{eqn:condition1zero}, \eqref{eqn:condition2mindirect} and \eqref{eqn:condition2mincross} with equality.
\label{lemma:lower_compress}
\end{lemma}
\begin{IEEEproof}[Proof of converse]
To prove Equation \eqref{eqn:condition2mindirect}, note that
\begin{eqnarray}
H_q(W_i^*|Z_m) = H_q(W_i^*,Z_m|Z_m) \ge H_q(W_i|Z_m).
\end{eqnarray}
But, $H_q(W_i|Z_m)$ is the number of subfiles of $W_i$ unknown to user $m$, multiplied by the size of one subfile, which is given by 
\begin{eqnarray}
H_q(W_i|Z_m) &=&\frac{\prod_{j=1}^{i-1}{K-r_{j+1} \choose r_j - r_{j+1}} {K-1\choose r_i} \prod_{j =i+1 }^{N}{r_{j-1} \choose r_j}}{\prod_{j=1}^{i-1}{K-r_{j+1} \choose r_j - r_{j+1}} {K\choose r_i} \prod_{j =i+1 }^{N}{r_{j-1} \choose r_j}}F\nonumber\\
 &=& \frac{{K -1 \choose r_i}}{{K \choose r_i}}F. 
\end{eqnarray}
 To prove Equation \eqref{eqn:condition2mincross}, let us concentrate on one arbitrary pair $(m,\ell)$ where $m\in \Omega_i$ and $\ell\in[K]\backslash\Omega_{i}$. 
\begin{align}
H_q(W_i^*|Z_\ell)&\ge H_q(W_i^*|Z_\ell,Z_m)= H_q(W_i^*,Z_m|Z_\ell,Z_m)\nonumber\\
&\stackrel{(a)}{\ge} H_q(W_i|Z_\ell,Z_m) \stackrel{(b)}{=} \frac{{K-2 \choose r_i}}{{K\choose r_i}}F,
\end{align}
where $(a)$ follows from Equation \eqref{eqn:condition1zero} and $(b)$ is due to our placement strategy. To see why $(b)$ holds, note that $H_q(W_i|Z_\ell,Z_m)$ is the number of subfiles of $W_i$ unknown to both users $\ell$ and $m$, multiplied by the size of one subfile. This is given by 
\begin{align}
H_q(W_i|Z_\ell,Z_m) &=\frac{\prod_{j=1}^{i-1}{K-r_{j+1} \choose r_j - r_{j+1}} {K-2\choose r_i} \prod_{j =i+1 }^{N}{r_{j-1} \choose r_j}}{\prod_{j=1}^{i-1}{K-r_{j+1} \choose r_j - r_{j+1}} {K\choose r_i} \prod_{j =i+1 }^{N}{r_{j-1} \choose r_j}}F \nonumber\\
&= \frac{{K -2 \choose r_i}}{{K \choose r_i}}F. 
\end{align}          
\end{IEEEproof}   
Most of this section will be dedicated to constructing $W_1^*$ and $W_2^*$ that satisfy the achievability part of Lemma \ref{lemma:lower_compress}. Once we have designed such $W_1^*$ and  $W_2^*$, we will construct a delivery message that helps all the users decode both.  

Let $(\rho_1,\rho_2)$ be an arbitrary pair of sets such that $\rho_2\subseteq\rho_1\subseteq \Omega_1$ and $s_i \stackrel{\bigtriangleup}{=} |\rho_i|\le r_i$ for $i\in[2]$. Let ${W}_{2,\rho_1,\rho_2}$ be a column vector whose elements are the subfiles of $W_2$ of the form $W_{2,\rho_1\cup x_1,\rho_2\cup x_2}$ for all $x_2\subseteq x_1\subseteq \Omega_2$. The order of the elements in ${W}_{2,\rho_1,\rho_2}$ is immaterial, as long as it is known to the users. This vector has $\kappa_2(s_1,s_2)$ elements where
\begin{eqnarray}
\kappa_2(s_1,s_2) &=& {K_2\choose r_2 - s_2}{K_2 - (r_2 - s_2) \choose r_1 - s_1 - (r_2 - s_2)}.
\end{eqnarray}
However, note that each user in $\Omega_2$ knows all but $\theta_2(s_1,s_2)$ elements of ${W}_{2,\rho_1,\rho_2}$ where
\begin{align}
\theta_2(s_1,s_2) &= {K_2 - 1\choose r_2 - s_2}{K_2 - (r_2 - s_2) \choose r_1 - s_1 - (r_2 - s_2)}\nonumber \\
&= \frac{K_2 - (r_2 -s_2)}{K_2}\kappa_2(s_1,s_2).
\end{align}
 From the perspective of a user in $\Omega_1$, the story is entirely different. He either knows the entire $W_{2,\rho_1,\rho_2}$ (if his index is in the set $\rho_2$) or he does not know anything about $W_{2,\rho_1,\rho_2}$. We shall encode the vector $W_{2,\rho_1,\rho_2}$ of length $\kappa_2(s_1,s_2)$ as a new vector $W^*_{2,\rho_1,\rho_2}$ of length $\theta_2(s_1,s_2)$ in such a way that decoding  $W^*_{2,\rho_1,\rho_2}$ enables each user in $\Omega_2$ to decode $W_{2,\rho_1,\rho_2}$. By doing so, we are simultaneously {\it aligning} the subfiles of $W_2$ which are unknown to the users in $\Omega_1$ to the extent possible.   
     

Let ${\cal C}_{2,s_1,s_2}$ be an arbitrary MDS matrix with $\theta_2(s_1,s_2)$ rows and $\kappa_2(s_1,s_2)$ columns. We know that if we remove any of $\kappa_2(s_1,s_2) - \theta_2(s_1,s_2)$ columns of ${\cal C}_{2,s_1,s_2}$, the resulting square matrix is invertible. Define 

\begin{eqnarray}
{W}^*_{2,\rho_1,\rho_2} = {\cal C}_{2,s_1,s_2}{W}_{2,\rho_1,\rho_2}.
\end{eqnarray}
Let $W^*_2$ be a vertical concatenation of all the vectors $W^*_{\rho_1,\rho_2}$ for all $(\rho_1,\rho_2)$. Let us calculate the length of the vector ${W}^*_2$.
\begin{align}
 &\mbox{length}({W}^*_2) = \sum_{{\rho_2\subseteq\rho_1\subseteq \Omega_1}}\theta_2(|\rho_1|,|\rho_2|)\nonumber\\
&= \sum_{s_1,s_2}   {K_1 \choose s_2}{K_1 - s_2\choose s_1 - s_2}\theta_2(s_1,s_2)\nonumber\\
&= \sum_{s_1,s_2}  {K_1 \choose s_2}{K_2 - 1\choose r_2 - s_2}{K_1- s_2\choose s_1-s_2}{K_2 - (r_2 - s_2) \choose (r_1 - r_2)  - (s_1 - s_2)}\nonumber\\
&= \sum_{s_2}   {K_1 \choose s_2}{K_2 - 1\choose r_2 - s_2}\sum_{s_3}{K_1- s_2 \choose s_3}{K_2 - (r_2 - s_2) \choose (r_1 - r_2)  - s_3}\nonumber\\
&\stackrel{(a)}{=} {K-1\choose r_2}{K- r_2\choose r_1 -r_2}=S\frac{{K-1\choose r_2}{K- r_2\choose r_1 -r_2}}{{K\choose r_2}{K- r_2\choose r_1 -r_2}}= S\frac{{K-1\choose r_2}}{{K\choose r_2}}.
\label{eqn:compressed_length}
\end{align}
where we have defined $s_3 = s_1 - s_2$, and $(a)$ follows from applying the Vandermonde identity to each summation. {We can also compute the number of subfiles in $W_2^*$ which are unknown to a user $j\in\Omega_1$ as
\begin{align}
&e_j= \sum_{\substack{{\rho_2\subseteq\rho_1\subseteq \Omega_1}\\{j\not\in \rho_2}}}\theta_2(|\rho_1|,|\rho_2|)\nonumber\\
&= \sum_{{\rho_2\subseteq\rho_1\subseteq \Omega_1}}   {K_1 -1 \choose s_2}{K_1 - s_2\choose s_1 - s_2}\theta_2(s_1,s_2)\nonumber\\
&= \sum_{s_1,s_2}   {K_1 -1 \choose s_2}{K_2 - 1\choose r_2 - s_2}{K_1- s_2\choose s_1-s_2}{K_2 - r_2 +s_2 \choose r_1 - r_2  - s_1 + s_2}\nonumber\\
&= \sum_{s_2}   {K_1 -1 \choose s_2}{K_2 - 1\choose r_2 - s_2}\sum_{s_3}{K_1- s_2 \choose s_3}{K_2 - r_2 + s_2 \choose r_1 - r_2  - s_3}\nonumber\\
&= {K-2\choose r_2}{K- r_2\choose r_1 -r_2} = S\frac{{K-2\choose r_2}{K- r_2\choose r_1 -r_2}}{{K\choose r_2}{K- r_2\choose r_1 -r_2}}= S\frac{{K-2\choose r_2}}{{K\choose r_2}}.
\label{eqn:compressed_unknown}
\end{align}
It is not difficult to see that $H_q(W_2^*|Z_j) = e_j\frac{F}{S}$ which matches the lower-bound presented in Lemma \ref{lemma:lower_compress}, Equation \eqref{eqn:condition2mincross}. Furthermore, $H_q(W_2^*|Z_m)$ for $m\in\Omega_2$ is upper-bounded by length$(W_2^*)\frac{F}{S}$ which matches Equation \eqref{eqn:condition2mindirect}.} Quite similarly, by reversing the roles of the two files in the description above, one can find a column vector ${W}^*_1$ of length $S\frac{{K - 1\choose r_1}}{{K\choose r_1}}$ {with $S\frac{{K - 2\choose r_1}}{{K\choose r_1}}$ subfiles unknown to any user in $\Omega_2$. This also serves as a proof for the achievability part of Lemma \ref{lemma:lower_compress}.
Now that we defined $W_1^*$ and $W_2^*$, it is left to construct a delivery message that enables all the users in $[K]$ to decode both. To accomplish this, we construct the delivery message as 
\begin{eqnarray}
X_{\bf d} = {\cal C}[{W}_1^*;{W}_2^*],
\end{eqnarray}
where ${\cal C}$ represent an MDS matrix with $S\max\left\{ \frac{{K-1\choose r_2}}{{K\choose r_2}} + \frac{{K-2\choose r_1}}{{K\choose r_1}} ,  \frac{{K-1\choose r_1}}{{K\choose r_1}} + \frac{{K-2\choose r_2}}{{K\choose r_2}} \right\}$  rows and $S\left(\frac{{K -1\choose r_1}}{{K \choose r_1}} + \frac{{K -1\choose r_2}}{{K \choose r_2}}\right)$ columns. Note that the number of rows of ${\cal C}$ is chosen to be the maximum number of subfiles of $[W_1^*;W_2^*]$ unknown to any user in $[K]$. The resulting delivery rate is}
\begin{eqnarray}
R = \max\left\{ \frac{{K-1\choose r_2}}{{K\choose r_2}} + \frac{{K-2\choose r_1}}{{K\choose r_1}} ,  \frac{{K-1\choose r_1}}{{K\choose r_1}} + \frac{{K-2\choose r_2}}{{K\choose r_2}} \right\}.
\label{eqn:another_R_1041}
 \end{eqnarray}
The delivery strategy has been summarized in Algorithm \ref{Alg:delivery}.

\begin{algorithm}
\caption[caption]{The Delivery Strategy for $N=2$ files and $K$ users}
\begin{algorithmic}[1]
\Statex {{\bf Input:} $(W_1,W_2), (r_1,r_2), {\bf d}, K$}
 \Statex {\bf Output: }{The delivery message $X_{\bf d}$.}
 \Statex{}
 \State  $\Omega_1 = \{i\in[K]| d_i = W_1\}$ and $\Omega_2= [K]\backslash \Omega_1$.
 \State  $K_i = |\Omega_i|$  for $i\in\{1,2\}$.
 \For{$ i\in [2]$}
 \If {$\Omega_i = [K]$}
 \State  $R = \frac{K - r_i}{K}$.
 \State Let ${\cal C}$ be an $SR$ by $S$ MDS matrix.
 \State Return $X_{\bf d} = {\cal C}W_i$.
\EndIf
\EndFor
\For {$i\in[2]$}
\For{ $s_1 \in[0:\min\{r_1,|\Omega_{[2]\backslash\{i\}}|\}]$}
\For{ $s_2 \in[0:\min\{r_2,s_1\}]$}
\State  $\kappa_i(s_1,s_2)={K_i\choose r_2 - s_2}{K_i - (r_2 - s_2) \choose r_1 - s_1 - (r_2 - s_2)}.$
\State  $\theta_i(s_1,s_2) = \frac{K_i - (r_i-s_i)}{K_i}\kappa_i(s_1,s_2).$
\State Let ${\cal C}_{i,s_1,s_2}$ be a  $\theta_i(s_1,s_2) $ by $ \kappa_i(s_1,s_2)$ MDS 
\Statex $\;\;\;\hspace{1.3cm}$matrix.
\For {$\rho_1 \subseteq \Omega_{[2]\backslash\{i\}}$ s.t. $|\rho_1 | = s_1$}
\For {$\rho_2 \subseteq \rho_1$ s.t. $|\rho_2 | = s_2$}
\State Let $W_{i,\rho_1,\rho_2}$ be a vertical concatenation
\Statex $\;\;\;\hspace{2.3cm}$of the subfiles $\left\{W_{i,{\rho_1\cup x_1},{\rho_2\cup x_2}}| x_2\subseteq \right.$
\Statex$\;\;\;\hspace{2.3cm}\left.x_1 \subseteq \Omega_i\right\}$.
\State  ${W}^*_{i,\rho_1,\rho_2} = {\cal C}_{i,s_1,s_2}{W}_{i,\rho_1,\rho_2}.$
\EndFor
\EndFor
\EndFor
\EndFor
\EndFor
\State   $R = \max\left\{ \frac{{K-1\choose r_2}}{{K\choose r_2}} + \frac{{K-2\choose r_1}}{{K\choose r_1}} ,  \frac{{K-1\choose r_1}}{{K\choose r_1}} + \frac{{K-2\choose r_2}}{{K\choose r_2}} \right\}$.
\State Let  ${\cal C}$ be a $SR$ by $S\left(\frac{{K -1\choose r_1}}{{K \choose r_1}} + \frac{{K -1\choose r_2}}{{K \choose r_2}}\right)$ MDS  matrix.
\For {$i\in [2]$}
\State Let $W_i^*$ be a vertical concatenation of the vectors $\left\{W^*_{i,\rho_1,\rho_2}| \rho_2\subseteq \rho_1\subseteq \Omega_{[2]\backslash \{i\}}\right\}$.
\EndFor
\State Return $X_{\bf d} = {\cal C} [W_1^*;W_2^*]$. 

\end{algorithmic}
\label{Alg:delivery}
\end{algorithm}

 \subsection{Correctness}
 In this section we prove the correctness of Algorithm \ref{Alg:delivery} by establishing that upon receiving $X_{\bf d}$ every user will be able to recover his requested file. The decoding process for each user is done in two phases reminiscent of a peeling algorithm. In the first phase, each user decodes both $W_1^*$ and $W_2^*$. In the second phase, each user $i$ discards $W^*_{\bar{d_i}}$ where $\bar{d_i}$  is the index of the file that has not been requested by user $i$. He then proceeds to decode $W_{d_i}$ using only $W^*_{i}$ and the side information stored in his cache.
 \subsubsection*{ Decoding Step 1}
 First let us show that after receiving $X_{\bf d}$, every user can recover the entire $[W_1^*;W_2^*]$. Remember that ${\cal C}$ is an MDS  matrix with $SR$  rows and $S\left(\frac{{K -1\choose r_1}}{{K \choose r_1}} + \frac{{K -1\choose r_2}}{{K \choose r_2}}\right)$ columns. Let $\gamma_i$ represent the set of columns of ${\cal C}$ corresponding to the elements of $[W_1^*;W_2^*]$ which user $i$ already knows from the side information available in his cache. Let $ {\cal C}_{\bar{\gamma}_i}$ represent the submatrix of ${\cal C}$ obtained by removing the columns indexed in $\gamma_i$. If this matrix is square (or overdetermined), user $i$ will be able to invert it and recover $[W_1^*;W_2^*]$. Therefore, we need to prove that $|\gamma_i|\ge S\left(\frac{{K -1\choose r_1}}{{K \choose r_1}} + \frac{{K -1\choose r_2}}{{K \choose r_2}}\right) -SR$. {This inequality directly follows from Equations \eqref{eqn:compressed_length}, \eqref{eqn:compressed_unknown} and \eqref{eqn:another_R_1041}: 
 \begin{align}
 |\gamma_i| &= S \frac{{K-1 \choose r_{\bar{d_i}} }}{{K\choose r_{\bar{d_i}}}}- S\frac{{K-2 \choose r_{\bar{d_i}}}}{{K\choose r_{\bar{d_i}}}}\nonumber\\
 &\ge  S\min\left\{\frac{{K-1 \choose r_1 }}{{K\choose r_1}}- \frac{{K-2 \choose r_1 }}{{K\choose r_1}}, \frac{{K-1 \choose r_2 }}{{K\choose r_2}}- \frac{{K-2 \choose r_2 }}{{K\choose r_2}}\right\}\nonumber\\
 & = S\left(\frac{{K -1\choose r_1}}{{K \choose r_1}} + \frac{{K -1\choose r_2}}{{K \choose r_2}}\right) -SR.
 \end{align}}
 \subsubsection*{ Decoding Step 2}
Now we show that if a user in $\Omega_1$ has the entire $W_1^*$, he can decode it for $W_1$. Similarly, if a user in $\Omega_2$ has the entire $W_2^*$, he can recover $W_2$. It should however be noted that users in $\Omega_{d_i}$ will not be able to recover $W_{\bar{d}_i}$.\\
The proof idea is very similar to Step 1. Remember that $W_1^* = \{W^*_{1,\rho_1,\rho_2}| \rho_2\subseteq\rho_1\subseteq \Omega_2\}$. We will show that once a user in $\Omega_1$ has access to $W^*_{1,\rho_1,\rho_2}$ he will be able to decode $\{W_{1,\rho_1\cup x_1, \rho_2\cup x_2}| x_2\subseteq x_1\subseteq \Omega_1\}$ for all $\rho_2\subseteq\rho_1\subseteq\Omega_2$. To see why, note that $W^*_{1,\rho_1,\rho_2} = {\cal C}_{1,s_1,s_2}W_{1,\rho_1,\rho_2}$ where $s_i = |\rho_i|$. The matrix  $ {\cal C}_{1,s_1,s_2}$ is an MDS matrix with $\theta_1(s_1,s_2)$ rows and $\kappa_1(s_1,s_2)$ columns. The number of subfiles of   $W_{1,\rho_1,\rho_2}$ unknown to user $i\in\Omega_1$ is equal to $\theta_1(s_1,s_2)$.  User $i$ can thus discard the remaining $\kappa(s_1,s_2) - \theta_1(s_1,s_2)$ columns of $ {\cal C}_{1,s_1,s_2}$ and invert the resulting square matrix in order to recover his unknowns. 

\subsection{Expected Achievable Rate}
To summarize, we characterized the delivery strategy for every choice of $(M_1,M_2)$ of the form $M_i = \frac{r_i}{K}$ with $r_i \in [0:K]$, and a non-trivial request vector. Two questions are left to be addressed. First, what if $M_iK$ is not an integer, and second, for a fixed total cache size of $M$, what are the optimal values of $M_1$ and $M_2$? To answer the first question, we observe that the lower convex envelope of all the points $(M_1,M_2,R)$ with $M_iK\in \mathbb{Z}$ is achievable by simply performing memory-sharing among such points. If $(M_1K,M_2K)$ is not a pair of integers, we rely on this memory-sharing strategy to find an achievability scheme. We postpone the second question to Section \ref{sec:binary_search}, once we have a better understanding of the optimal memory-sharing strategy for arbitrary $(M_1,M_2)$. For now, we write the achievable expected delivery rate as the minimum over all possible choices of $(M_1,M_2)$ that satisfy $M_1 + M_2 = M$.

\begin{theorem}
Consider the coded caching problem with $2$ files $W_1$ and $W_2$, and $K$ users each equipped with a cache of size $M$. Denote the probability of requesting file $W_i$ by $p_i$ where $p_1 + p_2 = 1$.  Then the following expected delivery rate is achievable. 
\begin{align}
\bar{R}(M) &= \min_{\substack{{t_1,t_2}\\{t_1 + t_2 = KM}}} \Big(\frac{K - t_1}{K}p_1^K  + \frac{K - t_2}{K}p_2^K\nonumber \\
&+(1 - p_1^K - p_2^K){\cal L}_{{\bf r}\rightarrow {\bf t}} \max ((R_1(r_1,r_2), R_2(r_1,r_2)) \Big) 
\label{eqn:achi}
\end{align}
where 
\begin{align}
R_1(r_1,r_2) &= \frac{{K - 1 \choose r_1}}{{K \choose r_1}} +  \frac{{K - 2 \choose r_2}}{{K \choose r_2}},\nonumber\\
R_2(r_1,r_2) &= \frac{{K - 2 \choose r_1}}{{K \choose r_1}} +  \frac{{K - 1 \choose r_2}}{{K \choose r_2}}.
\end{align}
for all $(r_1,r_2)\in \mathbb{Z}^2$ s.t. $0\le r_1, r_2 \le K$.
\label{thm:achievability}
\end{theorem}
 
 \begin{IEEEproof}
 Based on the proof of correctness of Algorithm \ref{Alg:delivery}, we know that if we allocate a cache of size $M_i = r_i/K$ to file $i$ for $i\in[2]$, and if both files are requested, then we can achieve a delivery rate of $R_{r_1,r_2} = \max\{R_1(r_1,r_2),R_2(r_1,r_2)\}$. If only file $i$  has been requested, then we can easily achieve a delivery rate of $\frac{K-r_i}{K}$. By performing memory-sharing among all such points $(r_1,r_2)$, we are able to achieve the lower convex envelope of all the points $((r_1,r_2), \bar{R}_{r_1,r_2})$ where     $\bar{R}_{r_1,r_2} = \frac{K-r_1}{K}p_1^K + \frac{K-t_2}{K}p_2^K + (1-p_1^K-p_2^K)R_{r_1,r_2}$, and  $0\le r_1,r_2\le K$ and $r_i\in \mathbb{Z}$. 
 By restricting this lower convex envelope to the plane which yields a cache of size $M$, we can characterize the achievable expected rate, $\bar{R}_{t_1,t_2}$, for all $(t_1,t_2)$ s.t. $t_1/K + t_2/K = M$ and $0\le t_1, t_2\le K$, $(t_1,t_2) \in \mathbb{R}^2$. We can then choose the pair $(t_1,t_2)$ for which the rate $\bar{R}_{t_1,t_2}$ is minimized.
\end{IEEEproof}

\begin{rem}
In the special case of $r_1 = r_2$, our joint placement and delivery strategy achieves the same delivery rate as in \cite{yu2017exact}. This is because $\frac{{K - 1 \choose r}}{{K \choose r}} +  \frac{{K - 2 \choose r}}{{K \choose r}}= \frac{{K \choose r+1} - {K-2 \choose r+1}}{{K\choose r}}$. Although the two placement strategies become equivalent when $r_1 = r_2$, the delivery strategies remain distinct. In other words, Algorithm \ref{Alg:delivery} offers an alternative delivery strategy for the uniform caching of two files, which can be of independent interest.  
\end{rem}

\begin{rem}
The key of our proposed delivery strategy is to construct, for each requested file $W_i,$ a compressed description $W_i^*$ that will be decoded by {\it every}  user, even those who did not request $W_i.$
One of the main results of our paper is that for the case of two users, this leads to optimal performance.
Unfortunately, the same basic strategy fails to attain the lower bound developed below in Section \ref{sec:converse}.
Nonetheless, we conjecture that this lower bound can indeed be attained using our general placement scheme together with an improved delivery strategy.
However, the delivery strategy must fundamentally rely on
the alignment of the undesired messages at each user, in a more subtle way than through the computation of $W_i^*$. In particular, such alignments must occur across the subfiles of multiple undesired files, not just one.  
\end{rem}

\section{Converse Bound for Uncoded Placement}
\label{sec:converse}
In this section we will derive a converse bound for the expected delivery rate under uncoded placement for arbitrary $K$, $N$ and ${\bf p}$. For a request vector ${\bf d}$, we define $Range({\bf d})$ as the set of indices of the files that are requested at least once in ${\bf d}$. That is, $Range({\bf d}) = \{i\in[N]\; |\; \exists j\in[K] \mbox{ s.t. } d_j = i\}$. {To prove our converse bound, we will follow in the footsteps of Lemma 2 in \cite{yu2017exact} which provides an uncoded converse bound for uniform file popularities. To start with, fix a request vector ${\bf d}$ and let ${\cal N} = Range({\bf d})$ and let $u_i$ be the index of an arbitrary user such that $d_{u_i} = W_i$, and let ${\cal U} = \{u_i|i\in{\cal N}\}$.}

The general idea behind the proof in \cite{yu2017exact} is to construct a virtual user whose cache contains a subset of the symbols stored by the users in ${\cal U}$. This is done in such a way that the virtual user can recover all the files $\{W_i| i\in{\cal N}\}$ after receiving the delivery message $X_{\bf d}$. For instance, let us fix a bijective function $\pi: [|{\cal N}|]\rightarrow {\cal N}$. The virtual user can store the entire cache of user  ${u_{\pi(1)}}$, but since $(X_{\bf d},Z_{u_{\pi(1)}})$ enables him to decode $W_{{\pi(1)}}$, he will only cache the symbols in $Z_{u_{\pi(2)}}$ which do not belong to $W_{\pi(1)}$. Similarly, he can discard the symbols in $Z_{u_{\pi(3)}}$ which belong to either $W_{\pi(1)}$ or $W_{\pi(2)}$, and so on. The converse bound is simply $H_q(\{W_i| i\in{\cal N}\} |Z)$ where $Z$ is the cache of the virtual user.

This converse bound depends on the particular request vector ${\bf d}$ and the choice of the ``leaders" ${\cal U}$. To remove these dependencies, one can take the average of the converse bound over all possible request vectors that share the same $Range({\bf d}) ={\cal N}$  
as well as all possible choices of the leaders. Finally, note that the converse bound also depends on $\pi(\cdot)$ which indicates in which order different files indexed in ${\cal N}$ are processed by the virtual user. We create one virtual user for each $\pi$. Since every such virtual user must be able to recover all the files $ \{W_i|i\in{\cal N}\}$, the overall converse bound will be the maximum of the $|{\cal N}|!$ bounds obtained in this fashion.

\begin{theorem}
Consider the problem of coded caching with $N$ files with probabilities $(p_1,\dots,p_N)$ and $K$ users such that each user has a cache of size $M$. The expected delivery rate under uncoded placement must satisfy

\begin{align}
\bar{R}\ge \min_{\substack{{{\bf t}}\\{\sum t_i = MK}\\{0\le t_i\le K}}}\left[\sum_{{\cal N}\subseteq [N]}\sum_{\substack{{\bf d}\in [N]^K\\Range({\bf d}) = {\cal N}}}\prod_{i = 1}^K p_{d_i} \max_{\pi: [|{\cal N}|]\rightarrow {\cal N}}R_\pi({\bf t},{\cal N})\right]
\label{eqn:converse}
\end{align}
where the maximum is taken over all bijections $\pi:[|{\cal N}|]\rightarrow{\cal N}$, and 
\begin{align}
R_\pi({\bf t},{\cal N}) &= \sum_{i\in  [|{\cal N}|]} \left[(1 - t_{\pi(i)}+  \lfloor t_{\pi(i)} \rfloor )  \frac{{K-i\choose \lfloor t_{\pi(i)} \rfloor }}{{K\choose \lfloor t_{\pi(i)} \rfloor }}\right. \nonumber\\
 &+ \left. (t_{\pi(i)} -  \lfloor t_{\pi(i)} \rfloor ) \frac{{K-i\choose \lfloor t_{\pi(i)} \rfloor + 1 }}{{K\choose \lfloor t_{\pi(i)}\rfloor + 1}} \right].
\end{align}
\label{thm:converse}
\end{theorem}
\begin{IEEEproof}
The proof closely follows that of Lemma 2 in \cite{yu2017exact} with a few minor but important differences.  To start with, we assume that all the users together have dedicated a total (normalized) cache of size $t_i$ to file $W_i$ where $0\le t_i\le K$. Since each user has a cache of size $M$, we must have $\sum_{i=1}^N t_i = KM$.  As can be seen in the statement of Theorem \ref{thm:converse}, the converse bound for a particular request vector ${\bf d}$, only depends on ${\bf d}$ through ${\cal N}$, the set of indices of the files that have been requested at least once. For a fixed request vector, we also define $\Omega_i\subseteq [K], i\in {\cal N}$ as the set of indices of the users who have requested file $W_i$. For $i\in {\cal N}$, let $u_i\in \Omega_i $ be the index of an arbitrary user who has requested file $W_{i}$, and let ${\cal U } = \{u_i|i\in{\cal N}\}$. Suppose an auxiliary user has access to the entire cache of user $u_{\pi(i)}$ except for the symbols which belong to the files within $\{W_{{\pi(\ell)}}| \ell\in[|{\cal N}|], \ell < i\}$, for all $i\in [|{\cal N}|]$. Provided that this auxiliary user has received $X_{\bf d}$, he must be able to recover all the files within $\{W_{i}| i\in {\cal N}\}$. For this to be feasible, the delivery rate must satisfy \cite{yu2017exact}

\begin{align}
&R({\bf t}, {\cal N}) \ge \nonumber\\
&\frac{1}{F}\sum_{i\in [|{\cal N}|]} \sum_{j = 1}^F {\bf 1}\Big({\cal K}_{{\pi(i)},j}\cap \{u_{\pi(\ell)}| \ell \in [|{\cal N}|], \ell \le i\} = \emptyset\Big),
\end{align}
where ${\cal K}_{{\pi(i)},j}$ represents the subset of the users that have cached the $j$'th symbol of file $W_{{\pi(i)}}$. We take the average of the expression above over all request vectors ${\bf d}$ that have the same $Range({\bf d}) = {\cal N}$ and over all possible choices of the set ${\cal U}$. We obtain
\begin{eqnarray}
R({\bf t}, {\cal N}) \ge \frac{1}{F}\sum_{i \in [|{\cal N}|]} \sum_{j = 1}^F \frac{{K - |{\cal K}_{{\pi(i)},j}|\choose i}}{{K\choose i}}.
\end{eqnarray}

\noindent Similarly, we can build a new virtual user for every possible bijection $\pi : [|{\cal N}|]\rightarrow {\cal N}$.  Each virtual user, gives us a new converse bound. Therefore, we have
\begin{eqnarray}
R( {\bf t},{\cal N}) \ge \frac{1}{F}\max_{\pi :[|{\cal N}|]\rightarrow {\cal N}}\sum_{i \in [|{\cal N}|]} \sum_{j = 1}^F \frac{{K - |{\cal K}_{{\pi(i)},j}|\choose i}}{{K\choose i}}.
\end{eqnarray}

\noindent Let $a_{n,i}$ represent the number of symbols of file $W_{i}$ cached by exactly $n$ users, normalized by $F$. We can write 
\begin{align}
R( {\bf t},{\cal N}) &\ge \max_{\pi :[|{\cal N}|]\rightarrow {\cal N}}\sum_{i \in[|{\cal N}|]} \sum_{n = 0}^K \frac{{K - n \choose i}}{{K\choose i}}a_{n,\pi(i)} \nonumber\\
&=\max_{\pi : [|{\cal N}|]\rightarrow {\cal N}} \sum_{i \in [|{\cal N}|]} \sum_{n = 0}^K \frac{{K - i \choose n}}{{K\choose n}}a_{n,\pi(i)}.
\label{eqn:sum_permutation_lower}
\end{align}

\noindent For any $i$, consider the sequence $c_{n,i} = \frac{{K-i\choose n}}{{K \choose n}}, n\in[0:K]$ where ${a \choose b} = 0$ if $b>a$. Let $g_i:
\mathbb{R}\rightarrow \mathbb{R}$ be the continuous piecewise linear function whose corner points are $c_{n,i}$. In other words,  
\begin{align}
g_i(x) = \begin{cases}
(1 - x + \lfloor x \rfloor) c_{
\lfloor x \rfloor,i} + (x - \lfloor x \rfloor)c_{\lfloor x \rfloor+1,i} \\
\hspace{3.5cm}\mbox{ if } \;\; \lfloor x \rfloor \in[0:K-1],\\
\\
(1-x)c_{0,i} + xc_{1,i}
\hspace{0.8cm}\mbox{ if }  \;\; \lfloor x \rfloor <0,\\
\\
(K-x)c_{K-1,i} + (x-K+1)c_{K,i}\\ 
\hspace{3.5cm}\mbox{ if }  \;\; \lfloor x \rfloor >K-1.
\end{cases}
\end{align}
Note that $g_i(x)$ is a convex function for any $i\in [|{\cal N}|]$ . Furthermore, the sequence $a_{n,i}$, $n\in[0:K]$ satisfies $\sum_{n = 0}^K a_{n,i} =1$ and $a_{n,i}\ge 0$. Therefore, by Jensen's inequality we have
\begin{align}
\sum_{n = 0}^{K} a_{n,i}c_{n,i} = \sum_{n = 0}^{K} a_{n,i}g_i(n) \ge g_i(\sum_{n = 0}^{K} na_{n,i}). 
\end{align}
But note that $\sum_{n = 0}^{K} na_{n,i} = t_i$. As a result,
\begin{align}
&\sum_{n = 0}^{K} a_{n,i}c_{n,i} \ge g_i (t_i)= (1 - t_i + \lfloor t_i \rfloor)c_{\lfloor t_i \rfloor,i} \nonumber\\
&+ (t_i - \lfloor t_i \rfloor)c_{\lfloor t_i \rfloor+1,i}\nonumber\\
&= (1 - t_i + \lfloor t_i \rfloor) \frac{{K-i \choose \lfloor t_i \rfloor}}{{K \choose \lfloor t_i \rfloor}} + (t_i - \lfloor t_i \rfloor)\frac{{K-i \choose \lfloor t_i \rfloor +1}}{{K \choose \lfloor t_i \rfloor + 1}}. 
\end{align}
Based on this, we can continue to bound Equation \eqref{eqn:sum_permutation_lower} as 

\begin{align}
& R( {\bf t},{\cal N})\ge\nonumber\\
&\max_{\pi : [|{\cal N}|]\rightarrow {\cal N}}\sum_{i \in [|{\cal N}|]} \left[(1 - t_{\pi(i)}+  \lfloor t_{\pi(i)} \rfloor )  \frac{{K-i\choose \lfloor t_{\pi(i)} \rfloor }}{{K\choose \lfloor t_{\pi(i)} \rfloor }} \right.\nonumber\\
&\left.+ (t_{\pi(i)} -  \lfloor t_{\pi(i)} \rfloor ) \frac{{K-i\choose \lfloor t_{\pi(i)} \rfloor + 1 }}{{K\choose \lfloor t_{\pi(i)}\rfloor + 1}} \right].
 \end{align}
 Taking the expected value of this expression over all ${\cal N}$ and the minimum of the resulting expression over all possible $(t_1,\dots,t_N)$ provides the desired lower bound.
 \end{IEEEproof}

\begin{rem}
The minimization problem in Theorem \ref{thm:converse} can be solved with standard convex optimization tools thanks to the fact that the right hand side of Equation \eqref{eqn:converse} as well as the minimization constraints are convex in ${\bf t}$. To establish this fact, one only needs to show that 
\begin{eqnarray}
J(x) =(1 - x+  \lfloor x \rfloor )  \frac{{K-i\choose \lfloor x \rfloor }}{{K\choose \lfloor x\rfloor }} + (x -  \lfloor x \rfloor ) \frac{{K-i\choose \lfloor x \rfloor + 1 }}{{K\choose \lfloor x\rfloor + 1}}
\end{eqnarray}
is convex in $x$. But this expression is piece-wise linear in $x$. So, it is sufficient to prove that the slopes of the consecutive segments of $J(x)$ increase by $x$. Or, in other words,
\begin{eqnarray}
\frac{{K-i\choose \lfloor x \rfloor + 1 }}{{K\choose \lfloor x\rfloor + 1}} - \frac{{K-i\choose \lfloor x \rfloor }}{{K\choose \lfloor x\rfloor }}  \le \frac{{K-i\choose \lfloor x \rfloor + 2 }}{{K\choose \lfloor x\rfloor + 2}} - \frac{{K-i\choose \lfloor x \rfloor +1}}{{K\choose \lfloor x\rfloor+1 }}.
\end{eqnarray}
This fact can be proven via elementary manipulations and is omitted for conciseness.
\end{rem}

\section{Optimality Result for $N  =2 $}
\label{sec:optimality}

In this section we prove that for the special case of $N = 2$, the converse bound provided by Equation \eqref{eqn:converse} is tight. {Our proof of optimality also sheds light on the points which contribute to the lower convex envelope at each $(M_1,M_2)$ in Equation \eqref{eqn:achi}. As it turns out, it is always sufficient to look at the vicinity of the point $(M_1,M_2)$, and perform memory-sharing among points of the form $(r_1,r_2)$ where $r_i \in \{\lfloor M_i K\rfloor, \lceil M_iK\rceil \}$. We start with a useful observation and then present a corollary of Theorem \ref{thm:achievability}.} 

\begin{proposition}
Suppose $K,r_1,r_2$ are three positive integers such that $0\le r_2 < r_1 \le K$. We have
\begin{eqnarray}
\frac{{K-1 \choose r_1 }}{{K \choose r_1}} +  \frac{{K-2 \choose r_2}}{{K \choose r_2}} \ge \frac{{K-2 \choose r_1 }}{{K \choose r_1}} +  \frac{{K-1 \choose r_2}}{{K \choose r_2}}
\label{eqn:whichdirection}
\end{eqnarray}
if and only if $r_1 + r_2 \le K$. 
\end{proposition}

\begin{IEEEproof}
Define $A_1 = \frac{{K-2 \choose r_1 }}{{K \choose r_1}} +  \frac{{K-1 \choose r_2}}{{K \choose r_2}} $ and $A_2 = \frac{{K-1 \choose r_1 }}{{K \choose r_1}} +  \frac{{K-2 \choose r_2}}{{K \choose r_2}} $. Each $A_i$ can be computed as
\begin{eqnarray}
{A_i} = \frac{(K-1)(2K-r_1-r_2) - r_i(K-r_i)}{K(K-1)}.
\end{eqnarray}
Therefore,
\begin{align}
A_2 - A_1 &= \frac{r_1(K-r_1) - r_2(K-r_2)}{K(K-1)}\nonumber\\
 &= \frac{(r_1-r_2)(K-r_1-r_2)}{K(K-1)}.
\end{align}
Given that $r_1 > r_2$, we have $A_2 -A_1 \ge 0$ if and only if $ r_1 + r_2 \le K$.
\end{IEEEproof}
\begin{corollary}
For the caching problem with $K$ users, two files with probabilities $p_1$, $p_2$ and cache size $M$, the following expected delivery rate is achievable for any $(t_1,t_2)\in \mathbb{R}^{2}$ that satisfies $t_1 + t_2 = MK$.
\begin{align}
\bar{R}_{t_1,t_2} &= p_1^K \frac{K-t_1}{K} + p_2^K\frac{K-t_2}{K}\nonumber\\
&+ (1 -p_1^K-p_2^K)\max(R_1,R_2),
\label{eqn:achionly3}
\end{align}
where
\begin{align}
R_1 &=(1 + \lfloor t_1\rfloor -t_1)\frac{{K-1 \choose \lfloor t_1\rfloor}}{{K \choose \lfloor t_1\rfloor}} + (t_1 - \lfloor t_1\rfloor)\frac{{K-1 \choose \lfloor t_1\rfloor + 1}}{{K \choose \lfloor t_1\rfloor + 1}}\nonumber\\
 &+ (1 + \lfloor t_2\rfloor -t_2)\frac{{K-2 \choose \lfloor t_2\rfloor}}{{K \choose \lfloor t_2\rfloor}} + (t_2 - \lfloor t_2\rfloor)\frac{{K-2 \choose \lfloor t_2\rfloor + 1}}{{K \choose \lfloor t_2\rfloor + 1}}, \\
R_2 &=(1 + \lfloor t_1\rfloor -t_1)\frac{{K-2 \choose \lfloor t_1\rfloor}}{{K \choose \lfloor t_1\rfloor}} + (t_1 - \lfloor t_1\rfloor)\frac{{K-2 \choose \lfloor t_1\rfloor + 1}}{{K \choose \lfloor t_1\rfloor + 1}} \nonumber\\
&+ (1 + \lfloor t_2\rfloor -t_2)\frac{{K-1 \choose \lfloor t_2\rfloor}}{{K \choose \lfloor t_2\rfloor}} + (t_2 - \lfloor t_2\rfloor)\frac{{K-1 \choose \lfloor t_2\rfloor + 1}}{{K \choose \lfloor t_2\rfloor + 1}}.
\label{eqn:tworates}
\end{align}
\label{lm:only3}
\end{corollary}
\begin{IEEEproof}
We distinguish between two regimes.

\noindent {\bf Regime 1}. $t_1 - \lfloor t_1 \rfloor +t_2 - \lfloor t_2 \rfloor \ge 1$. We will perform memory sharing between three points $(r_1,r_2) \in T$ where $T = \{ (\lfloor t_1\rfloor , \lfloor t_2 \rfloor +1), (\lfloor t_1\rfloor + 1 , \lfloor t_2 \rfloor ),(\lfloor t_1\rfloor + 1 , \lfloor t_2 \rfloor +1)\}$. The coefficients that we use for memory-sharing are respectively $\theta_1,\theta_2,\theta_3$  where $\theta_1 = 1 - t_1 + \lfloor t_1 \rfloor$ , $\theta_2 =1 - t_2 + \lfloor t_2\rfloor$ and $\theta_3 = 1 -\theta_1  - \theta_2 =  t_1 - \lfloor t_1 \rfloor +t_2- \lfloor t_2\rfloor - 1$. The amount of cache dedicated to file $W_1$ is thus $\frac{\lfloor t_1\rfloor \theta_1 + (1+\lfloor t_1\rfloor) \theta_2 + (\lfloor t_1\rfloor + 1 ) \theta_3}{K} = \frac{t_1}{K}$.  Similarly, the amount of cache dedicated to file $W_2$ is $\frac{t_2}{K}$. \\
 If only one file $i$ is requested, we can trivially achieve $R = \frac{K - t_i}{K}$. 
Let us assume both files have been requested. There are two possibilities. Either $\lfloor t_1 \rfloor + \lfloor t_2 \rfloor + 2 \le K$ or $\lfloor t_1 \rfloor + \lfloor t_2 \rfloor + 1 \ge K$. In the first case, we know that Inequality  \eqref{eqn:whichdirection} holds for every $(r_1,r_2) \in T$. As a result, the following delivery rate can be achieved 

\begin{align}
R &= (1 -  t_1 + \lfloor t_1\rfloor)\frac{{K-1 \choose \lfloor t_1\rfloor}}{{K \choose \lfloor t_1\rfloor}} + (1 - t_1 + \lfloor t_1\rfloor)\frac{{K-2\choose \lfloor t_2\rfloor + 1}}{{K \choose \lfloor t_2\rfloor + 1}}\nonumber\\
 &+ (1 - t_2 +\lfloor t_2\rfloor)\frac{{K-1 \choose \lfloor t_1\rfloor+1}}{{K \choose \lfloor t_1\rfloor + 1}} + (1 - t_2 + \lfloor t_2\rfloor)\frac{{K-2 \choose \lfloor t_2\rfloor }}{{K \choose \lfloor t_2\rfloor }}\nonumber\\
  &+ (t_1 - \lfloor t_1 \rfloor+t_2- \lfloor t_2\rfloor - 1)\frac{{K-1 \choose \lfloor t_1\rfloor+1}}{{K \choose \lfloor t_1\rfloor+1}}\nonumber \\
&+( t_1 - \lfloor t_1 \rfloor +t_2- \lfloor t_2  \rfloor-1)\frac{{K-2 \choose \lfloor t_2\rfloor+ 1 }}{{K \choose \lfloor t_2\rfloor + 1}}\nonumber\\
  &=(1 + \lfloor t_1\rfloor -t_1)\frac{{K-1 \choose \lfloor t_1\rfloor}}{{K \choose \lfloor t_1\rfloor}} + (t_1 - \lfloor t_1\rfloor)\frac{{K-1 \choose \lfloor t_1\rfloor + 1}}{{K \choose \lfloor t_1\rfloor + 1}}\nonumber\\ 
& + (1 + \lfloor t_2\rfloor -t_2)\frac{{K-2 \choose \lfloor t_2\rfloor}}{{K \choose \lfloor t_2\rfloor}} + (t_2 - \lfloor t_2\rfloor)\frac{{K-2 \choose \lfloor t_2\rfloor + 1}}{{K \choose \lfloor t_2\rfloor + 1}}\nonumber\\
& = R_1.
\end{align}
In the second case, the direction of Inequality \eqref{eqn:whichdirection} is reversed for all $(r_1,r_2)\in T$. In this case, the following delivery rate can be achieved
\begin{align}
R &= (1 -  t_1 + \lfloor t_1\rfloor)\frac{{K-2 \choose \lfloor t_1\rfloor}}{{K \choose \lfloor t_1\rfloor}} + (1 - t_1 + \lfloor t_1\rfloor)\frac{{K-1\choose \lfloor t_2\rfloor + 1}}{{K \choose \lfloor t_2\rfloor + 1}}\nonumber\\
 &+ (1 - t_2 +\lfloor t_2\rfloor)\frac{{K-2 \choose \lfloor t_1\rfloor+1}}{{K \choose \lfloor t_1\rfloor + 1}} + (1 - t_2 + \lfloor t_2\rfloor)\frac{{K-1 \choose \lfloor t_2\rfloor }}{{K \choose \lfloor t_2\rfloor }}\nonumber\\
  &+ (t_1 - \lfloor t_1 \rfloor+t_2- \lfloor t_2\rfloor - 1)\frac{{K-2 \choose \lfloor t_1\rfloor+1}}{{K \choose \lfloor t_1\rfloor+1}}\nonumber \\
&+ ( t_1 - \lfloor t_1 \rfloor +t_2- \lfloor t_2  \rfloor-1)\frac{{K-1 \choose \lfloor t_2\rfloor+ 1 }}{{K \choose \lfloor t_2\rfloor + 1}}\nonumber\\
  &=(1 + \lfloor t_1\rfloor -t_1)\frac{{K-2 \choose \lfloor t_1\rfloor}}{{K \choose \lfloor t_1\rfloor}} + (t_1 - \lfloor t_1\rfloor)\frac{{K-2 \choose \lfloor t_1\rfloor + 1}}{{K \choose \lfloor t_1\rfloor + 1}}\nonumber\\  
&+(1 + \lfloor t_2\rfloor -t_2)\frac{{K-1 \choose \lfloor t_2\rfloor}}{{K \choose \lfloor t_2\rfloor}} + (t_2 - \lfloor t_2\rfloor)\frac{{K-1\choose \lfloor t_2\rfloor + 1}}{{K \choose \lfloor t_2\rfloor + 1}}\nonumber\\
& = R_2.
\end{align}
Therefore, we are able to achieve $\max(R_1,R_2)$. By taking the expectation over all possible request vectors, we achieve $\bar{R}_{t_1,t_2}$ as in Equation \eqref{eqn:achionly3}.  
 
\noindent {\bf Regime 2}. $t_1 - \lfloor t_1 \rfloor +t_2 - \lfloor t_2 \rfloor \le 1$. In this case we choose our set $T = \{ (\lfloor t_1\rfloor + 1 , \lfloor t_2 \rfloor ),(\lfloor t_1\rfloor , \lfloor t_2 \rfloor +1),(\lfloor t_1\rfloor , \lfloor t_2 \rfloor )\}$ and our coefficients $\theta_1 = t_1 - \lfloor t_1 \rfloor$ , $\theta_2 =t_2 - \lfloor t_2\rfloor$ and $\theta_3 =  1 - t_1 + \lfloor t_1 \rfloor -t_2+ \lfloor t_2\rfloor $. Again we perform memory-sharing between the three points in $T$ with the given coefficients. This allows us to achieve a delivery rate of $\frac{K - t_i}{K}$ if only file $W_i$ is requested. If both are requested, we can achieve $R = \max\{R_1,R_2\}$. By taking the expectation over all request vectors, we find the same delivery rate as in Equation \eqref{eqn:achionly3}.
\end{IEEEproof}
It is easy to see that the achievable rate characterized by Corollary \ref{lm:only3} lies on our converse bound in Equation \eqref{eqn:converse}. This implies that at any cache allocation point $(t_1,t_2)$, there are only three points $(r_1,r_2)$ that contribute to the lower convex envelope. We first check whether $t_1 - \lfloor t_1 \rfloor +t_2 - \lfloor t_2 \rfloor \ge 1$. If this inequality holds, then we perform memory sharing among the three points $ \{ (\lfloor t_1\rfloor , \lfloor t_2 \rfloor +1), (\lfloor t_1\rfloor + 1 , \lfloor t_2 \rfloor ),(\lfloor t_1\rfloor + 1 , \lfloor t_2 \rfloor +1)\}$. Otherwise, we perform memory-sharing among $\{ (\lfloor t_1\rfloor + 1 , \lfloor t_2 \rfloor ),(\lfloor t_1\rfloor , \lfloor t_2 \rfloor +1), (\lfloor t_1\rfloor , \lfloor t_2 \rfloor )\}$. Based on this observation, we can summarize our joint placement and delivery strategy as in Algorithm \ref{Alg:joint}. The next theorem follows immediately from Corollary \ref{lm:only3} and Theorem \ref{thm:converse}.

\begin{algorithm}
\caption[caption]{The joint placement-delivery strategy for $N=2$ and arbitrary $M,K,{\bf p}$}
\begin{algorithmic}[1]
\Statex {{\bf Input:} $W_1,W_2, M, K,{\bf p}$}
 \Statex {\bf Output: }{The cache contents $(Z_1,\dots,Z_K)$ and all the delivery messages $\{X_{\bf d}| {\bf d}\in [N]^K\}$.}
 \Statex{}
 \State  ${\bf t} = (t_1,t_2) = \argmin \bar{R}_{t_1,t_2}$ where $\bar{R}_{t_1,t_2}$ is given by Equation \eqref{eqn:achionly3}.
 \State $r_1 = \lfloor t_1\rfloor$, $r_2  = \lfloor t_2 \rfloor$.
 \Statex{}
\Statex {\bf Placement}
 \If { $t_1 - r_1 + t_2 - r_2 \ge 1$}
 \State  $\theta_0 = 0, \theta_1 = 1-t_1 + r_1, \theta_2 = 1 - t_2 + r_2$, 
\Statex $\;\;\;\;\hspace{1pt}\;\theta_3 = 1 - \theta_1 -\theta_2$. 
 \State  $Q_1 = (r_1,r_2+1), Q_2 = (r_1+1, r_2 ),$ 
\Statex $\;\;\;\hspace{1pt}\;\;Q_3 = (r_1+1,r_2 + 1)$.
 \Else 
 \State  $\theta_0 = 0, \theta_1 = t_1- r_1, \theta_2 =  t_2 - r_2$, $\theta_3 = 1 - \theta_1 -\theta_2$. 
 \State  $Q_1 = (r_1+1,r_2), Q_2 = (r_1, r_2+1 ), Q_3 = (r_1,r_2 )$.
 \EndIf
 \State  $P_j= \sum_{i=0}^j \theta_i$ for $j\in[0:3]$.
 \State  $W_{i}^{j} ={W_{i}|_{[P_{j-1}F+1:P_jF]}}$ for $i\in[2]$, $j\in[3]$ where ${W_{i}|_A}$ refers to the symbols of $W_i$ indexed in in the set $A$. 
 \State  $(Z_1^i,\dots,Z_K^i) = $ output of Algorithm \ref{Alg:placement} applied on $((W^i_{1},W^i_2),Q_i, K)$ for $i\in[3]$.
\State  $Z_j = {(Z^{1}_j,Z^{2}_j,Z^{3}_j)} $ for $j\in [K]$. 
\Statex{}
\Statex {\bf Delivery}
\For {all request vectors ${\bf d}$}
\State  $X^{(i)}_{\bf d}$ = output of Algorithm \ref{Alg:delivery} applied on $((W^i_1,W^i_2),Q_i,{\bf d},K)$ for $i\in[3]$.
\State  $X_{\bf d} = (X^{(1)}_{\bf d},X^{(2)}_{\bf d},X^{(3)}_{\bf d})$.
\EndFor
\State Return $((Z_1,\dots,Z_K),\{X_{\bf d}| {\bf d}\in [N]^K\})$.
\end{algorithmic}
\label{Alg:joint}
\end{algorithm}

\begin{theorem}
For the coded caching problem with $K$ users, two files with probabilities $p_1,p_2$ and cache size $M$,  the optimal expected delivery rate under uncoded placement  is
\begin{eqnarray}
\bar{R}^*= \min_{\substack{{0\le t_2\le t_1\le K}\\{t_1 + t_2 = KM}}}\bar{R}_{t_1,t_2},
\label{eqn:optimaleasy}
\end{eqnarray}
 where $\bar{R}_{t_1,t_2}$ is given by Equation \eqref{eqn:achionly3}. Furthermore, this can be achieved by the joint placement and delivery strategy in Algorithm \ref{Alg:joint}.
\label{thm:optimality}
\end{theorem}


\subsection{Finding the optimal memory allocation}
\label{sec:binary_search}

The delivery rate in Equation \eqref{eqn:achionly3} as a function of $(t_1,t_2)$ is convex. As a result, the optimal $(t_1,t_2)$, which is the solution to 
\begin{eqnarray}
(t_1^*,t_2^*) = \argmin_{\substack{{0\le t_2\le t_1\le K}\\{t_1 + t_2 = KM}}}\bar{R}_{t_1,t_2},
\label{eqn:argeasy}
\end{eqnarray}
  can be found by solving a convex optimization problem. However, note that our delivery rate is in fact a piece-wise linear function of $(t_1,t_2)$, the break points of which can be easily characterized. Based on the following theorem, we can find the optimal $(t_1,t_2)$  by simply performing binary search over a discrete set of feasible points.
  
\begin{theorem}
There exists a solution $(t_1^*,t_2^*)$ to Equation \eqref{eqn:argeasy} that satisfies $(t_1^*,t_2^*)\in {\cal P}$ and
\begin{eqnarray}
 m^+(t_1^*) &\ge& \frac{p_1^K - (1-p_1)^K }{K(1 - p_1^K - (1-p_1)^K)},\nonumber\\
 m^{-}(t_1^*) &\le& \frac{p_1^K - (1-p_1)^K }{K(1 - p_1^K - (1-p_1)^K)},
\label{eqn:bestm}
\end{eqnarray}
where we define
\begin{align}
m^+(t_1) &\stackrel{\bigtriangleup}{=}\begin{cases}
\frac{{K - 1\choose \lfloor t_1 \rfloor+ 1 }}{{K \choose \lfloor t_1 \rfloor+ 1 }} - 
\frac{{K - 1\choose \lfloor t_1\rfloor }}{{K \choose \lfloor t_1 \rfloor }} + 
\frac{{K - 2\choose \lceil KM - t_1 \rceil -1}}{{K \choose \lceil KM - t_1\rceil -1 }} - 
\frac{{K - 2\choose \lceil KM - t_1 \rceil }}{{K \choose \lceil KM - t_1\rceil }}
\\
\hspace{5.4cm} \mbox{ if }  M \le 1,\\
\\
\frac{{K - 2\choose \lfloor t_1 \rfloor+ 1 }}{{K \choose \lfloor t_1 \rfloor+1 }} - 
\frac{{K - 2\choose \lfloor t_1\rfloor }}{{K \choose \lfloor t_1 \rfloor }} + 
\frac{{K - 1\choose \lceil KM - t_1 \rceil -1}}{{K \choose \lceil KM - t_1\rceil -1 }} - 
\frac{{K - 1\choose \lceil KM - t_1 \rceil }}{{K \choose \lceil KM - t_1\rceil }}
\\
 \hspace{5.4cm}\mbox{ if } M> 1.
\end{cases} \nonumber
\end{align}
\begin{align}
m^-(t_1) &\stackrel{\bigtriangleup}{=}\begin{cases}
\frac{{K - 1\choose \lceil t_1 \rceil }}{{K \choose \lceil t_1 \rceil}} - 
\frac{{K - 1\choose \lceil t_1\rceil -1 }}{{K \choose \lceil t_1 \rceil -1 }} + 
\frac{{K - 2\choose \lfloor KM - t_1 \rfloor }}{{K \choose \lfloor KM - t_1\rfloor }} - 
\frac{{K - 2\choose \lfloor KM - t_1 \rfloor+ 1 }}{{K \choose \lfloor KM - t_1\rfloor + 1 }}\\
\hspace{3.6cm} \mbox{ if }  t_1 >t_2 \mbox{ and } M \le 1, \\
\\
\frac{{K - 2\choose \lceil t_1 \rceil }}{{K \choose \lceil t_1 \rceil}} - 
\frac{{K - 2\choose \lceil t_1\rceil -1 }}{{K \choose \lceil t_1 \rceil -1 }} + 
\frac{{K - 1\choose \lfloor KM - t_1 \rfloor }}{{K \choose \lfloor KM - t_1\rfloor }} - 
\frac{{K - 1\choose \lfloor KM - t_1 \rfloor+ 1 }}{{K \choose \lfloor KM - t_1 \rfloor + 1 }}\\
\hspace{3.6cm} \mbox{ if }  t_1 >t_2 \mbox{ and }  M> 1,\\
\\
-m^+(t_1) 
\hspace{4.05cm}\mbox{ if } t_1 = t_2.
\end{cases} \nonumber
\end{align}
\begin{align}
{\cal P} &\stackrel{\bigtriangleup}{=} \Big\{(t_1,t_2)\in \mathbb{R}^2| 0\le t_2\le t_1\le K,\;\; t_1 + t_2 = KM,\nonumber\\
&(t_1 - \lfloor t_1 \rfloor)(t_2 - \lfloor t_2 \rfloor)(t_1 - t_2) = 0\Big\}.
\end{align}
\label{thm:search}
\end{theorem}
\begin{IEEEproof}
As we limit our achievable delivery rate to a line $t_1 + t_2 = KM$, we obtain a piecewise linear and convex curve $\bar{R}(t_1)$. It can be readily seen from Equation \eqref{eqn:achionly3} that the break points of this curve occur when $t_1 \in \mathbb{Z}$ or $t_2 \in \mathbb{Z}$ or at the extreme point when $t_1 = t_2$. This establishes the choice of the feasible sets ${\cal P}$ in the statement of the theorem. \\
If $M\le 1$, we know that $R_1$ is the maximizer of Equation \eqref{eqn:achionly3}. We can thus rephrase the expected delivery rate as

\begin{align}
&\bar{R}_{t_1,t_2} = (1 -p_1^K-p_2^K)\left[(1 + \lfloor t_1\rfloor -t_1)\frac{{K-1 \choose \lfloor t_1\rfloor}}{{K \choose \lfloor t_1\rfloor}} \right.\nonumber\\
&+ (t_1 - \lfloor t_1\rfloor)\frac{{K-1 \choose \lfloor t_1\rfloor + 1}}{{K \choose \lfloor t_1\rfloor + 1}} + 
(1 + \lfloor t_2\rfloor -t_2)\frac{{K-2 \choose \lfloor t_2\rfloor}}{{K \choose \lfloor t_2\rfloor}} \nonumber\\
&\left.+ (t_2 - \lfloor t_2\rfloor)\frac{{K-2 \choose \lfloor t_2\rfloor + 1}}{{K \choose \lfloor t_2\rfloor + 1}}\right]+p_1^K \frac{K-t_1}{K} + p_2^K\frac{K-t_2}{K} \nonumber\\
&= t_1 \left[(1 -p_1^K-(1-p_1)^K)\left( \frac{{K - 1\choose \lfloor t_1 \rfloor+ 1 }}{{K \choose \lfloor t_1 \rfloor+ 1 }} - 
\frac{{K - 1\choose \lfloor t_1\rfloor }}{{K \choose \lfloor t_1 \rfloor }}  \right.\right.\nonumber\\
&+\left.\left. \frac{{K - 2\choose \lfloor KM - t_1 \rfloor }}{{K \choose \lfloor KM - t_1\rfloor }} - 
\frac{{K - 2\choose \lfloor KM - t_1 \rfloor+ 1 }}{{K \choose \lfloor KM - t_1\rfloor + 1 }} \right) + \frac{(1-p_1)^K - p_1 ^K}{K}\right] + c.
\end{align}

Define ${\cal Q} \stackrel{\bigtriangleup}{=} \left\{t_1\in \mathbb{R}| \exists t_2 \in \mathbb{R}\;\;  s.t. \;\;  (t_1,t_2)\in {\cal P}\right\}. $ As long as $t_1$ is in the open interval between two fixed consecutive members of ${\cal Q}$, the value of $c$ does not change. As a result, the expression above provides us with the slope of the line segment which connects two consecutive points in the piecewise linear function $\bar{R}(t_1)$. Our goal is to find the value of $t_1\in{\cal Q}$ such that the slope of this curve is non-negative at $t_1+ \epsilon$ and non-positive at $t_1 -\epsilon$.  This is given by Equation \eqref{eqn:bestm}. Note that we are using the identity $(1 + \lfloor a \rfloor  - a)g(\lfloor a \rfloor) + (a - \lfloor a \rfloor)g(\lfloor a \rfloor + 1)=(\lceil a \rceil  - a)g(\lceil a \rceil -1) + (a - \lceil a \rceil   + 1)g(\lceil a \rceil)$ to simplify the expressions for $m^+(t_1)$ and $m^-(t_1)$.
Similar analysis can be made if $M> 1$.
\end{IEEEproof}

{
\section{Numerical Results}
\label{sec:numerical}

\begin{figure}
\centering
\includegraphics[scale=0.3]{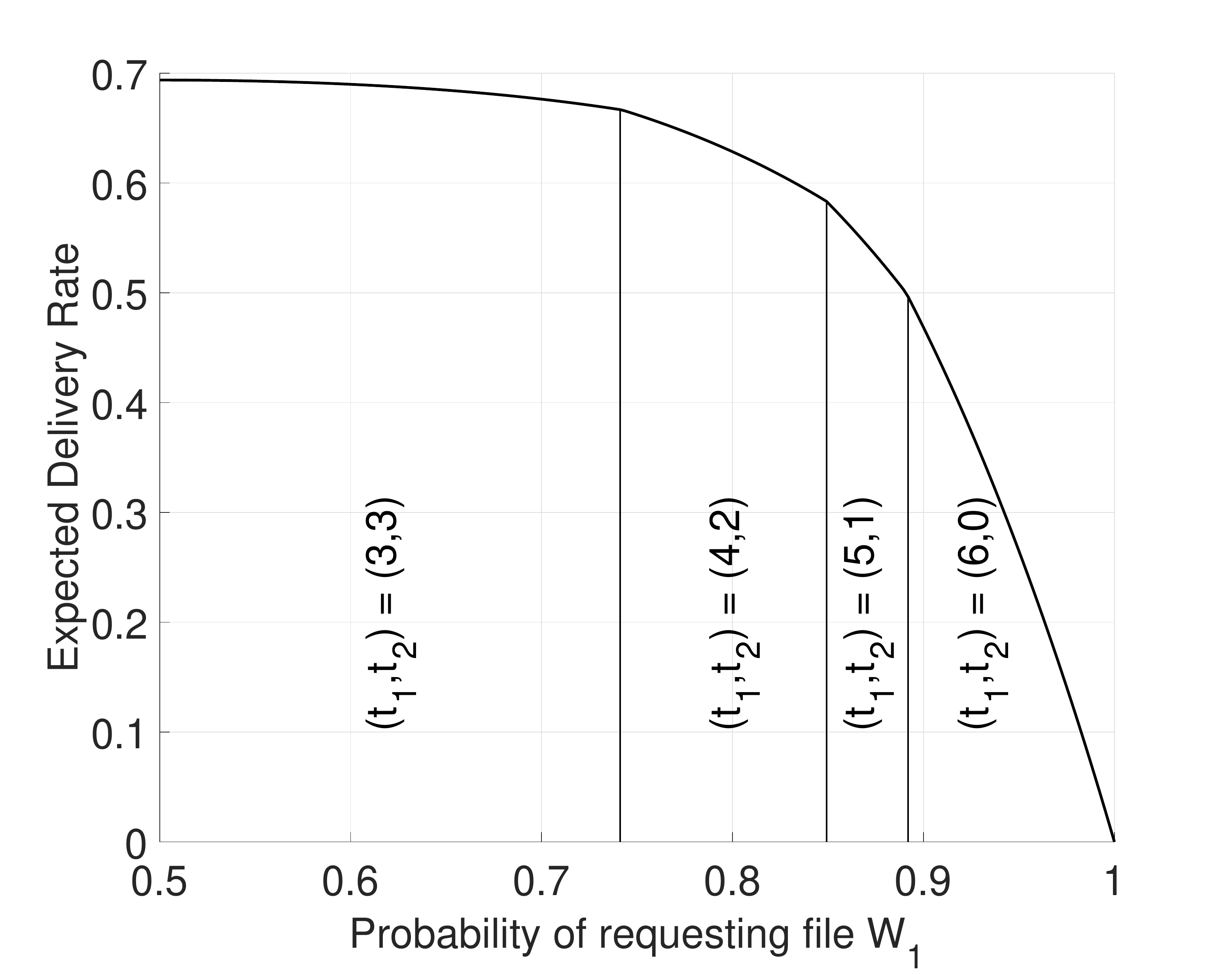}
\caption{{The optimal expected delivery rate (under uncoded placement) for the non-uniform caching problem with $K=6$, $N=2$ and $M=1$, versus the probability of requesting $W_1$.}}
\label{fig:vsProbs}
\end{figure}

\begin{figure}
\centering
\begin{subfigure}[b]{0.45\textwidth}
\includegraphics[width=\textwidth]{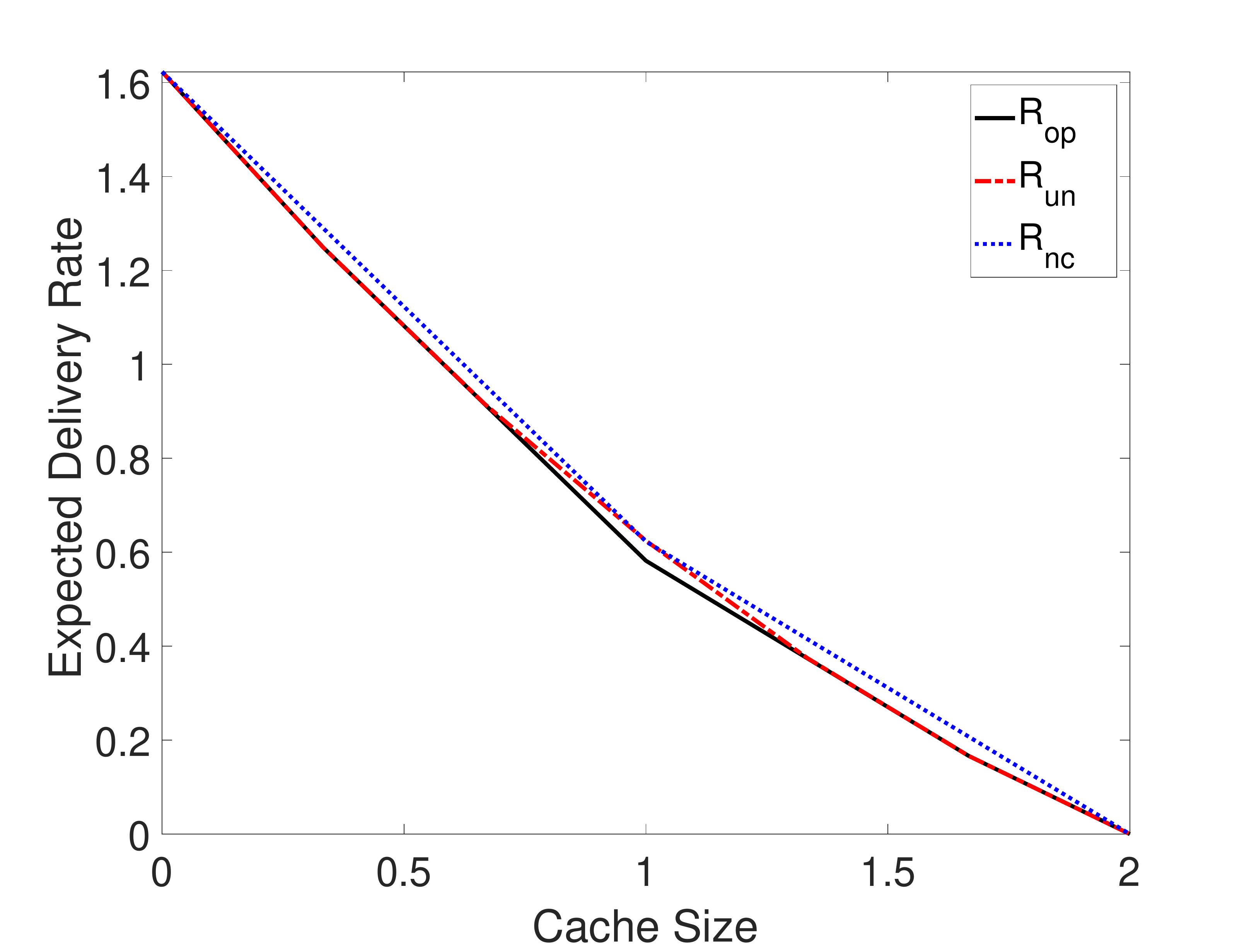}
\end{subfigure}
\begin{subfigure}[b]{0.45\textwidth}
\includegraphics[width=\textwidth]{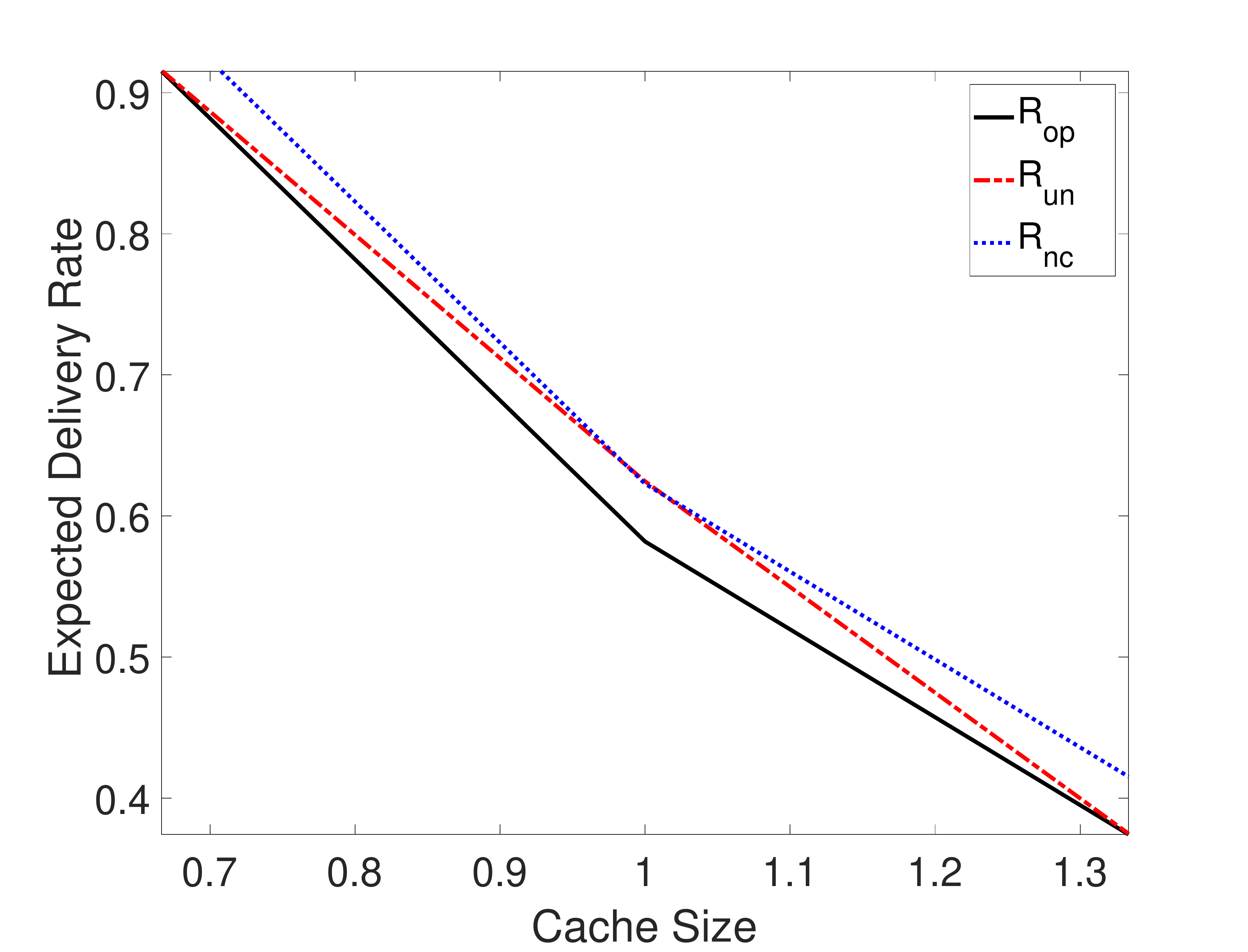}
\end{subfigure}
\caption{{{\bf top}: Comparison of the expected delivery rate for our scheme $R_{\mbox{op}}$ (optimal under uncoded placement), and the two possible grouping strategies: $R_{\mbox{un}}$ performs uniform caching and ignores the differences in the probabilties, whereas $R_{\mbox{nc}}$ ignores the coding opportunities between the two files.  The parameters are $K=6$, $N=2$ and $p_1 = 0.85$. {\bf bottom}: zoomed in on the vicinity of $M = 1$.}}
\label{fig:vsCache}
\end{figure}
In this section we provide a numerical analysis of our caching strategy and compare it with the literature. First, we fix $K = 6$, $M = 1$ and $N = 2$, and find the optimal expected delivery rate (under uncoded placement) as a function of the probability of the first file, using Theorem \ref{thm:optimality}. To accomplish this, we first have to find the optimal $(t_1,t_2) = (M_1K,M_2K)$ as a function of $p_1$ following Theorem \ref{thm:search}. This optimal expected delivery rate has been plotted in Figure \ref{fig:vsProbs}. A few valuable insights can be gained from this curve. Firstly, when the probabilities of the two files are close, the heuristic approach of applying uniform coded caching is indeed optimal. The range of probabilities for which this property holds ultimately depends on $K$ and $M$, but for our example is given by $|p_1 - p_2| < 0.48$. This is the region that has been marked by $(t_1,t_2) = (3,3)$ on the figure. Similarly, when one file is very popular (in this case $|p_1 - p_2| > 0.78$), it is optimal to allocate the entire cache to it, and ignore the other file in the placement phase. This region has been labeled as $(t_1,t_2) = (6,0)$.}

But perhaps the most interesting scenario is when the probabilities lie somewhere in between. Figure \ref{fig:vsProbs}  tells us that there is a range of probabilities (in this example $0.48 < |p_1 - p_2| < 0.78$) for which no memory-sharing strategy is optimal. For this range, one must rely on Algorithm \ref{Alg:joint} with non-trivial choices of $(t_1,t_2)$ to attain the optimal expected delivery rate. For this particular example, we must set $(t_1,t_2)= (4,2)$ for $0.48<|p_1 - p_2|<0.70 $ and $(t_1,t_2) = (5,1)$ for $0.70<|p_1 - p_2|< 0.78$.

Now, let us instead fix $p_1$ and find the optimal expected delivery  rate as a function of the cache size $M$. For $K=6$, we rely on the previous plot to choose $p_1 = 0.85$ in order to emphasize the scenario where grouping is strictly sub-optimal, at least at $M=1$. In Figure \ref{fig:vsCache} we have plotted the optimal expected delivery rate for this choice of $p_1$, and compared it to the two possible grouping strategies  \cite{niesen2017coded,ji2014average,ji2017order,zhang2018coded}: $R_{\mbox{un}}$ corresponds to the uniform caching which ignores the differences in the probabilties of the two files, whereas $R_{\mbox{nc}}$ is the delivery rate for a caching strategy that creates two groups each containnig one file, and ignores the coding opportunities between the two. The expression for $R_{\mbox{un}}$ can be given \cite{yu2017exact} by the  lower convex envelop of the points  
\begin{align}
R_{\mbox{un}} &= p_1^K({1-\frac{r}{K}}) + (1-p_1)^K(1-\frac{r}{K}) \nonumber\\
&+ (1-p_1^K - (1-p_1)^K)\left[\frac{{K \choose r+1} - {K-2 \choose r+1}}{{K \choose r}}\right],
\end{align}
where $r = \frac{KM}{2}\in \mathbb{N}$. As for $R_{\mbox{nc}}$, it is easy to see that if $M\le 1$, the best memory allocation is to assign the entire cache to $W_1$. If $M>1$, the remaining memory is given to file $W_2$. This results in a delivery rate of
\begin{align}
R_{\mbox{nc}} = \begin{cases}
(1-p_1^K)(2-M) & \mbox{ if } M>1,\\
-p_1^K - (1-p_1)^K(1-M) + 2-M & \mbox{ if } M \le 1. 
\end{cases}
\end{align} 
As visible in Figure \ref{fig:vsCache}, the most discrepancy between the grouping and optimal strategies occur around $M=1$, where the optimal expected delivery rate is $0.582$, whereas $R_{nc}\approx 0.623$ and $R_{un}\approx 0.625$, about 7 percent larger than the optimal rate.  On the other hand, at the extreme values of $M$, all three strategies are optimal. This is not very surprising: if $M=0$ or $M=2$, all three strategies are equivalent. It is therefore natural that in the vicinity of such extreme values there is no major difference in their performances.

\section{Concluding Remarks}
\label{sec:conclusion}
The majority of the existing literature on coded caching with non-uniform demands is focused on grouping strategies which can achieve constant additive or multiplicative gaps to the optimal expected delivery rate. This paper serves as a step towards the ambitious task of designing nonuniform coded caching strategies which are {\it optimal} under uncoded placement. Moreover, we believe that there is great potential to the multiset indexing extension of the uniform placement strategy proposed in this paper, as it can be readily applied to other combinatorial problems of heterogeneous nature. Our delivery strategy for the case of two files may also serve as a stepping stone for a closer investigation of the application of interference alignment in coded caching.

\section*{Acknowledgement}
The authors would like to thank the Associate Editor and the reviewers for their invaluable feedback.

\bibliographystyle{IEEEtran}
\bibliography{IEEEfull,nonuniform}

\begin{IEEEbiographynophoto}{Saeid Sahraei}
received his Ph.D. and M.S. in 2018 and 2013, both from the School of Computer and Communication Sciences, \'Ecole Polytechnique F\'ed\'erale de Lausannne (EPFL). Prior to that, he obtained his B.S. in Electrical Engineering from Sharif University of Technology in 2010. He is currently a postdoctoral scholar at the Department of Electrical Engineering, University of Southern California (USC). His research interests are in information theory, coding theory, combinatorial optimization and computational complexity.

Dr. Sahraei received the SNSF Early Postdoc.Mobility Fellowship in 2018 and the EPFL EDIC Doctoral Fellowship in 2013.
\end{IEEEbiographynophoto}

\begin{IEEEbiographynophoto}{Pierre Quinton}
 received a master's degree in Data Science from \'Ecole Polytechnique F\'ed\'erale de Lausanne (EPFL), Lausanne, Switzerland, in 2019. During this time he worked on information theory and coding with Prof. Michael Gastpar and Prof. Emre Telatar. He is currently pursuing his PhD degree at EPFL under the supervision of Prof. Emre Telatar in the Information Theory Laboratory within the School of Computer and Communication Sciences. His main research interests lie within the areas of information theory, probability theory, and statistics.
\end{IEEEbiographynophoto}

\begin{IEEEbiographynophoto}{Michael Gastpar}
(S'99--M'03--SM'14--F'17) received the Dipl. El.-Ing. degree from the Eidgen\"ossische Technische Hochschule (ETH), Z\"urich, Switzerland, in 1997, the M.S. degree in electrical engineering from the University of Illinois at Urbana-Champaign, Urbana, IL, USA, in 1999, and the 
Doctorat \`es Science degree from the Ecole Polytechnique F\'ed\'erale (EPFL), Lausanne, Switzerland, in 2002. He was also a student in engineering and philosophy at the Universities of Edinburgh and Lausanne. 

During the years 2003-2011, he was an Assistant and tenured Associate Professor in the Department of Electrical Engineering and Computer Sciences at the University of California, Berkeley. Since 2011, he has been a Professor in the School of Computer and Communication 
Sciences, Ecole Polytechnique F\'ed\'erale (EPFL), Lausanne, Switzerland. 
He was also a professor at Delft University of Technology, The Netherlands, and 
a researcher with the Mathematics of Communications Department, 
Bell Labs, Lucent Technologies, Murray Hill, NJ. 
His research interests are 
in network information theory and related coding and signal processing techniques, 
with applications to sensor networks and neuroscience. 

Dr. Gastpar received the IEEE Communications Society and Information Theory Society Joint Paper Award in 2013 and the EPFL Best Thesis Award in 2002. He was an Information Theory Society Distinguished Lecturer (2009-2011), an Associate Editor for Shannon Theory for the IEEE TRANSACTIONS ON INFORMATION THEORY (2008-2011), and he has served as Technical Program Committee Co-Chair for the 2010 International Symposium on Information Theory, Austin, TX. 
\end{IEEEbiographynophoto}

\end{document}